\documentclass[11pt, a4paper]{article}
\usepackage[utf8]{inputenc}
\usepackage{graphicx}
\graphicspath{ {images/} }
\usepackage{url}
\usepackage{color}
\usepackage{natbib}
\usepackage[left=2cm,right=2cm,top=2cm,bottom=2cm]{geometry}
\usepackage{setspace}
\usepackage{hyperref}
\usepackage{float}
\usepackage{amsmath}
\usepackage{enumitem}
\usepackage{todonotes}
\usepackage{multirow}
\usepackage{makecell}
\usepackage{threeparttable}
\usepackage{booktabs}
\usepackage{longtable}
\usepackage{caption}
\usepackage{subcaption}

\begin{document}
\sloppy

\begin{center}
\huge Going Beyond the In-/Out-Group Dichotomy: \\ Investigating Altruism towards Middle-Groups\footnote{
\textbf{Acknowledgements:} We thank Ming Dai, Leonie Loy and Patricia von Mellenthin for help in the
data collection; Tamas Olah for supporting the oTree implementation; Vinicius Ferraz, Wolf Gardian
and Thomas Pitz for enabling the data collection at Rhine-Waal University of Applied Sciences; Melina
Hühn and Guido Sprenger for complementing this research project through ethnographic interviews;
and Eugen Dimant, Christina Rott, Martin Strobel and participants at ESA Bologna 2022 for fruitful
discussions.
\textbf{Funding:} The study was funded by Heidelberg University.
\textbf{Institutional Review Board Certificate:}  https://gfew.de/ethik/RxfgxQ88 by the German As-
sociation for Experimental Economic Research e.V.
\textbf{Conflict of Interest:} The authors report no conflict of interest.}  \\
\end{center}
\normalsize

\noindent Leon Houf\textsuperscript{1} \& Christiane Schwieren\textsuperscript{2}   \\
\footnotesize\textsuperscript{\textbf{1}}Karlsruhe Institute of Technology (KIT), corresponding author: Leon.Houf@kit.edu\\
\footnotesize\textsuperscript{\textbf{2}}Heidelberg University\\
\normalsize
\today \\

\begin{abstract}
In-group favouritism and out-group hostility are well-documented, but real-world group settings rarely fit a simple dichotomy. Often, a ``middle-group'' shares some identity markers with the in-group without fully belonging to it. How do people treat such intermediate groups?
We address this question using a formal identity marker framework and a multi-lab online experiment ($N = 376$) with a private allocation task immune to reputation effects and demand characteristics. We find that a middle-group can be treated neutrally, i.e., distinct from both in-group favouritism and out-group hostility, but only under specific conditions. When the middle-group shares two identity markers with the in-group, including university affiliation, a complete hierarchy emerges: the in-group is favoured, the middle-group is treated according to the allocation rule, and the out-group bears the loss.  However, when the middle-group shares only one marker, it is treated indistinguishably from the out-group, regardless of which marker is shared. Furthermore, differentiated behaviour evolves only after participants have re-encountered the experimental instructions and group constellation in a second phase, not gradually across rounds.
These findings demonstrate that participants can perceive beyond binary in-/out-group categorisation, but such perception requires both sufficient identity overlap and repeated exposure to the group setting. Our identity marker framework provides a tool for systematically studying group relations in more complex settings.
\end{abstract}


\section{Introduction}

Humans are social creatures that tend to organise themselves into groups. This can take many forms, ranging from families and tribes to football clubs, shared flat communities, or fully online communities, such as those on Discord. In formal organisations and companies, groups or departments are established. Looking at this pattern of group formation, central questions in the social sciences are: How do groups coordinate and maintain internal order? And how do the members of a certain group act towards the members of their own, versus members of other groups?

One identified mechanism for behaviour in groups is in-group favouritism. In this mechanism, the in-group is treated favourably, creating long-term benefits for in-group members \citep{fu2012evolution}.
In addition to in-group favouritism, a distinct out-group is often treated unfavourably or even with hostility. This mechanism of out-group hostility is often found to stabilise (in-)group identity, yet is not necessarily the inverse of in-group favouritism \citep{allport1954nature, brewer1999psychology, dimant2023hate}. Notably, \cite{grigoryan2020perceived} provides experimental evidence that increased in-group favouritism need not be accompanied by out-group hostility. Even in groups with intense internal bonding, participants did not exhibit increased harm toward out-groups.
In-group favouritism and out-group hostility are studied in a variety of ways and contexts \citep{goette2012impact, rusch2014evolutionary, benistant2019unethical, li2020group, ciccarone2020rationale}, often confirmed by the minimal group paradigm \citep{tajfel1978social, chen2009group} and even in fully remote contexts \citep{amichai2005internet, janneck2013minimal}. Such patterns are often discussed as parochial altruism in contexts involving costly inter-group conflict  \citep{bernhard2006parochial, yamagishi2016parochial, pisor2024parochial}.

However, not every setting fits into a dichotomy of just an in-group and an out-group.
There are many scenarios with a group ``in the middle" of a clear and bounded in- and out-group, where the middle-group shares some identity markers with the in-group yet is clearly distinct from it. 
For instance, we can think of corporate workplaces, where a sales team forms a clear and bounded in-group in contrast to their competitors, the out-group. 
Here, other departments of the same company (e.g., finance) share characteristics and goals with the in-group. Yet, these departments are neither part of the in-group of the sales team nor an out-group of a competing company. 
Similarly, academic institutes or faculties of one particular discipline might regard interdisciplinary teams within their university as a middle-group compared to other disciplines as out-group.
Many more examples can be constructed where a simple dichotomy of in- and out-group falls short.

The effects of the regulating forces of in-group favouritism and out-group hostility on the middle-group are a priori unclear, and studies on similarities of groups, with third parties or multiple out-groups have yielded mixed results. \citet{grimm2017group} examine discrimination when facing multiple out-groups and find that relative similarity to the in-group affects treatment. \citet{kranton2020deconstructing} decompose behaviour in allocation tasks into ``groupy" and ``not-groupy" components, showing substantial individual heterogeneity. Relatedly, \cite{lee2021hidden} show that individuating information about personal preferences can reduce prosocial behaviour toward in-group members by weakening reliance on group membership labels, suggesting that not all identity-relevant information strengthens group-based treatment. \citet{attanasi2016social} show that social connectedness—a form of identity overlap—improves coordination on efficient outcomes. Empirically, even a single shared identity marker can elevate cooperation across group lines: \citet{chuah2014religion} show in a multicultural experiment that sharing a religious affiliation significantly increased trust and cooperation even between different ethnic groups. Recent studies confirm that greater overlap in group identities generally amplifies in-group bias \citep{hong2022multidimensional}. However, if group affiliations conflict, behaviour can deviate from the usual `us vs.\ them' pattern. \citet{hong2022multidimensional} even report cases of individuals favouring an out-group over their nominal in-group when two social identities clash. Therefore, compressing all such interactions into an in-/out-group dichotomy might conceal underlying behavioural factors that regulate groups.

While this body of work establishes that identity overlap matters, important questions remain. Most prior studies rely on stated preferences, hypothetical allocations, or strategic games with feedback. In such settings reputation effects, reciprocity, or demand characteristics may confound the measurement of group-based preferences. Furthermore, existing designs typically vary either the number of shared identities or the presence of multiple out-groups, but do not systematically distinguish which specific markers are shared or test whether marker type matters beyond marker count. Finally, few studies examine how group perceptions evolve within an experimental session.

Our study addresses these gaps. We introduce a formal \textit{identity marker framework} that allows systematic variation in both the number and type of shared markers between a middle-group and the in-group. We measure \textit{actual behaviour in a fully private, non-strategic setting}: participants allocate real resources by (potentially) breaking an unenforceable rule, with no possibility of observation, feedback, or reputation effects. This provides a cleaner measure of intrinsic group-based preferences than prior paradigms. We also examine \textit{temporal dynamics}, showing that participants initially treat middle-groups like out-groups but shift toward differentiated treatment after repeated exposure. Finally, we explicitly test whether a complete \textit{hierarchy of groups} (In $>$ Middle $>$ Out) can be statistically confirmed.

Relative to \citet{hong2022multidimensional}, who study how overlapping identities affect bias toward a fixed in-group, we focus on how participants treat a third party (the middle-group) whose overlap with the in-group is experimentally varied. Relative to \citet{grimm2017group}, who examine discrimination among multiple out-groups, we hold the out-group constant and vary the middle-group. Our contribution is thus to isolate the effect of partial identity overlap on a specific intermediate category, using a behavioural measure free of strategic confounds.

Specifically, every individual in our design belongs to only one group. The in-group is constructed such that all members share the same identity markers; the out-group differs on all markers; and the middle-group shares some but not all markers with the in-group. We hold the identity markers of each participant and group constant, varying only which groups are present in the allocation task. While this connects to the concept of social identity complexity \citep{roccas2002social}, our focus is not on individuals' perceptions of their own group identities, but on observed behaviour toward groups with varying identity overlap.

We investigate our central research question—how altruistic actions towards a middle-group compare to those directed at in- and out-groups—through a multi-lab, lab-in-zoom experiment. Participants are grouped by four criteria: arbitrary painting preferences based on the minimal group paradigm, real associations to a university outside the laboratory, presence within the same lab-in-zoom session, and random allocation to a group within a session. They perform a 60-round die-rolling task under complete privacy, in which they are given a rule to allocate tokens based on die outcomes. They can easily break these rules, as they cannot be observed or punished. Because participants cannot be observed and there is no feedback mechanism, we exclude potential motives such as reciprocity, social status concerns, or experimental demand effects. This allows us to operationalise rule-breaking behaviour as purely altruistic action to benefit members of specified groups.

Our results confirm the general finding of in-group favouritism, where the in-group is always treated favourably. Due to the zero-sum nature of our experimental game, the out-group consistently bears this loss. 
When a middle-group shares sufficient identity markers in both quantity and perceived quality with the in-group, and the experimental participants have already been repeatedly performing the task and have been presented with the instructions and the group-setting twice, then the middle-group is indeed treated neutrally based on the expected value of the die-roll allocation rule. Still, the in-group is treated favourably, illustrating a clear instance of group hierarchy that extends beyond a dichotomy. 
When a middle-group shares few identity markers with an in-group, it is treated similarly to an out-group, even when participants have repeatedly encountered the instructions and done the task.

Our findings have two main implications for understanding human behaviour across different group settings. 
First, we document that participants see beyond a simple in-/out-group dichotomy: When there is a middle-group that shares sufficient identity markers with the in-group, the in-group is treated favourably, the middle-group neutrally, and the out-group bears the loss. Yet when a middle-group does not show sufficient identity markers, it is treated like another out-group, i.e., following an apparent ‘us versus them’ attitude.  
Secondly, we find that participants change their behaviour over the two parts in our game, even when the task and group settings remain constant for the entire experiment duration. This shows that findings from experimental designs studying group behaviour in one shot or short learning phases should not be overgeneralised as they might only illustrate the short-term effects of an intervention, especially in more complex group settings.
Therefore, this paper demonstrates that research designs beyond the in-/out-group dichotomy can and should be employed more frequently.

The remainder of the paper is structured as follows: Section 2 outlines a framework for identity markers and group settings. Section 3 presents the design, and Section 4 presents the results of our experiment. Section 5 concludes.

\section{Group and Identity Marker Framework}\label{framework}
In real-world social contexts, group relationships rarely fall into a strict binary of only “us” versus “them.” Instead, individuals often recognise gradations of group membership - a spectrum ranging from a clear in-group through various intermediate affiliations to a definitive out-group. We introduce a formal framework of \textit{identity markers} to describe these group relations. This framework enables us to move beyond the in-/out-group dichotomy by defining an in-between, or \textit{middle}, group category for individuals who share some identity markers with the in-group but are not part of it.

\subsection{Identity Markers and Group Definitions}
We define a set of identity markers, $x_1, x_2, x_3, x_4$, as distinguishing characteristics that delineate group membership. Each marker $x_i$ represents a binary attribute: two individuals either share $x_i$ (denoted simply by $x_i$) or do not share it (denoted by $x'_i$). Conceptually, these markers are treated as independent dimensions of identity; however, in specific settings, some markers may be hierarchically related (nested).

An \textbf{in-group} ($I$) is characterised by a common set of identity markers. Formally, for a given reference individual (the “self”), their in-group can be represented as the vector of all shared markers: 
\[ 
    I = [\,x_1,\,x_2,\,x_3,\,\,x_4\,]\,,
\] 
meaning all group members share markers $x_1, x_2, x_3$ and $x_4$ with the self. This defines the prototypical “us.” 

In contrast, an \textbf{out-group} ($O$) is defined by a clear difference on all these salient identity markers. We denote the out-group as: 
\[ 
    O = [\,x'_1,\,x'_2,\,x'_3,\,x'_4\,]\,,
\] 
where $x'_i$ is the contrasting counterpart to marker $x_i$. Thus, members of $O$ share no identity markers with the in-group individual, representing the archetypal “them.”

\subsection{Middle-Groups: Neither In- nor Out-group }
Between these extremes lies the \textbf{middle group}. We use the term “middle” to denote a group that is neither an in-group nor an out-group, but rather somewhere in between. A middle group shares some but not all identity markers with the in-group. Formally, any group $M$ in this intermediate category can be represented as a mix of shared and unshared markers:
\[ 
    M = [\,m_1,\,m_2,\,m_3,\,m_4\,]\,,
\] 
where each $m_i \in \{x_i,\,x'_i\}$ and the pattern $(m_1, m_2, m_3, m_4)$ contains at least one $x_i$ (shared marker) and at least one $x'_j$ (differing marker). In other words, $M$ overlaps with $I$ on some markers and with $O$ on others. This captures the idea that $M$ is neither “us” nor “them” but an ambiguous middle category.

To illustrate, consider a simple example with three identity dimensions ($x_1, x_2, x_3$). The in-group $I = [x_1, x_2, x_3]$ shares all three markers. A potential middle-group might share two of these markers but differ on one, such as $M_{(1,2)} = [x_1, x_2, x'_3]$. A more distant middle-group may share only a single marker, such as $M_{(1)} = [x_1, x'_2, x'_3]$.

This conceptualisation allows us to position groups along a continuum between full identity overlap and complete difference, rather than forcing a binary classification. Further, this framework can accommodate a wide range of configurations, including cases where identity dimensions are hierarchically structured. For example, in some settings, sharing marker $x_4$ may imply sharing $x_3$ and $x_2$, while other markers may be independent.

The particular identity markers and group constellations used in our experiment will be introduced in Section \ref{design}.

\subsection{Group Placement on a Continuum}
Rather than a sharp dichotomy, the identity marker model suggests a continuum or hierarchy of group-relatedness based on the number and perceived importance of shared markers, with the clear anchors of in- and out-group. At one end lies the in-group ($I$), fully overlapping on all markers; at the other end lies the out-group ($O$), with no overlap. Between these are a gradation of partially overlapping groups. We can visualize this on a spectrum in table \ref{imo-line}, where $M_{(1,2)}$ (sharing two markers) is positioned closer to $I$, and $M_{(1)}$ (sharing one marker) is closer to $O$. The exact placement between $M_{1,2}$ and $M_{2,3}$, as well as $M_1$ and $M_2$ in relation to $I$ and $O$ is an empirical question of the employed markers. In our analysis, we later present the empirical distribution of the markers used in this study in Figure \ref{middle-part2}.

\begin{table}[H]
\centering
\begin{tabular}{c|ccccc|c}
&&&&&& \\
In-Group & \hspace{1.5cm} & Middle-Group & \hspace{1.5cm} & Middle-Group & \hspace{1.5cm} &  Out-Group\\ 
& \hspace{1.5cm} & (2 markers) & \hspace{1.5cm} & (1 marker) & \hspace{1.5cm} &  \\ \cline{1-1} \cline{3-3} \cline{5-5} \cline{7-7} 
&&&&&& \\
 &  & $M_{1,2}$ &  & $M_{1}$ &  &  \\
I & $\leftrightarrow$ &  & $\leftrightarrow$ &  & $\leftrightarrow$ & O \\
 &  & $M_{2,3}$ &  & $M_{2}$ &  &  \\
 &&&&&& \\
\end{tabular}
\caption{Group placement on continuum. The in-group and out-group are clearly bounded. The different middle-groups are in the space between. Their exact placement, i.e., closeness to in- or out-group is an empirical question given the exact markers in question.} \label{imo-line}
\end{table}

\subsection{Theoretical Expectations: Threshold versus Additive Effects}

How might identity markers combine to influence behaviour toward middle-groups? We consider two possibilities.

Under an \textit{additive model}, each shared marker independently increases perceived similarity, and the total effect equals the sum of individual marker effects. 

Under a \textit{threshold model}, a minimum level of identity overlap is required before a group is perceived as distinct from the out-group. Below this threshold, all non-in-groups are categorised as ``them"; above it, graded distinctions emerge. This predicts that some single markers, regardless their individual strength, may be insufficient to cross the threshold, while marker combinations may exceed it.

Our design allows us to distinguish these accounts. If markers operate additively, we should observe $M_U > M_P$ (or $M_P > M_U$) across all comparisons, reflecting the relative strength of each marker. If a threshold operates, we may observe $M_U \approx M_P$ (both below threshold) even while $M_{P,U} \neq M_{P,S}$ (differences emerge only above threshold).

The latter pattern is what we observe: one-marker middle-groups are treated indistinguishably from each other and from out-groups, while two-marker middle-groups show differentiated treatment depending on marker composition. This suggests that participants require multiple convergent identity signals before adjusting behaviour away from a binary in-/out-group categorisation.

\section{Experimental Design}\label{design}

In this study, we aim to identify social preferences towards experimental groups. To achieve this, we control the assignment of identity characteristics. This enables us to employ causal identification methods to address our research questions. 
We implement the study as no-deception, ``lab-in-zoom" and multi-lab study programmed in oTree \citep{chen2016otree}. The design, sample size and analysis follow the pre-registration (\url{https://www.socialscienceregistry.org/trials/9670}). Additional treatments (T3 and T5 in the nunmbering in this paper) are pre-registered here at (\url{https://aspredicted.org/kxwv-wqry.pdf}). In appendix \ref{pre-reg}, all pre-registered analysis and instructions are presented, and in the main text, the most relevant results and their insights are presented.

Our study sample consisted of $N=376$ participants from Heidelberg University (52\%) and Rhine-Waal University of Applied Sciences (48\%), Germany.
The sample consisted of individuals who were 56\% female, 42\% male, and 2\% non-binary, with an average age of 25.9. Participants represented more than 40 nationalities, primarily German (56\%) and Indian (11\%). The major fields of study were Economics (23\%) and Engineering (10\%). About 23\% were participating in an experiment for the first time, while 25\% had participated in more than five experiments.  
The first experimental sessions took place in July and August 2022. The second experimental sessions for treatments 3 and 5 took place in September 2025. On average, participants earned 10.15€ and spent 31 minutes on the experiment.

\subsection{Group Assignment and Identity Markers}

As outlined in section \ref{framework}, we conceptualise group identity in terms of shared \textbf{identity markers}. In our experimental design, we implement four concrete specific markers, denoted $x_1$ through $x_4$, which define the group affiliations:

\begin{itemize}
    \item \textbf{$x_1=x_P$ Painting Preference:} An arbitrary preference between a Klee versus a Kandinsky painting, following a minimal group paradigm (based on the procedure of \cite{chen2009group}). Participants choose their preferred artist as the first task in the experiment, creating two artificial groups.
    \item \textbf{$x_2=x_U$ University Affiliation:} A real-life institutional identity marker based on the university where the session is conducted (Heidelberg University vs. Rhine-Waal University of Applied Sciences). This represents a “lived reality” identity. University affiliation has been shown to generate measurable group identity effects in economic experiments \citep{chen2022competition}.
    \item \textbf{$x_3=x_S$ Shared Session:} An experimental-context marker indicating that individuals attended the \textbf{same lab-in-zoom-session} at the same time. While participation in the lab-in-zoom session is anonymous (no camera, microphone, group chat or names for participants), it may still serve as an important differentiating marker.
    \item \textbf{$x_4=x_G$ Shared Group Task:} A salient group marker indicating that individuals cooperated in the same 3-person group task. Within each session (shared $x_S$), participants who share $x_P$ (painting preference) are randomly divided into groups of three and complete a short task designed to foster a feeling of “group success.” Sharing $x_U$, $x_P$ and $x_S$ is therefore a necessary, but not sufficient, condition to be placed in the same group ($x_G$).  In this task, each participant is shown two paintings — one by Klee, one by Kandinsky — and must identify which painting corresponds to the artist they previously selected. This occurs over five rounds, with new paintings in each round. If, as a group, they correctly identify more paintings than they miss, all members receive the same monetary reward. While decisions are made individually and there is no communication between group members, the shared task and collective outcome aim to reinforce group salience. This setup adheres to the minimal group paradigm, in which group identity is triggered without actual interaction. The task was intentionally designed to be non-trivial but solvable: on average, individuals identified 4.5 out of 5 paintings correctly, and approximately 97\% of participants correctly identified at least three paintings. Accordingly, all but one group successfully completed the task, and all successful participants received the same positive feedback, indicating that their group had solved the task successfully before moving on to the main task. The one group that did not solve the group task was excluded from the analysis as pre-registered (\url{https://aspredicted.org/kxwv-wqry.pdf}).
\end{itemize}

Using these markers, we can formally represent the group categories in our experiment in terms of marker overlaps (see section \ref{framework}). The \textbf{in-group (I)} for a given participant is defined by possessing all four identity markers in common: 
\[I = [x_P,\;x_U,\;x_S,\;x_G]\,.\] 
That is, an in-group member shares the same painting preference, university, session, and group for the group task. In contrast, an \textbf{out-group (O)} member differs on all these dimensions: 
\[O = [x'_P,\;x'_U,\;x'_S,\;x'_G]\,.\] 
Participants are told that out-group members “chose the other painting” ($x'_P$), are from another university ($x'_U$), and are in a different session at a different time ($x'_S$), meaning they could not have participated in the same group task ($x'_G$). Thus, the out-group is operationalised as a group of distant strangers with no identity markers in common with the in-group.

Between these extremes, we introduce \textbf{middle-groups} that share some but not all markers with the in-group. In our design, we consider four types of middle-groups corresponding to five treatment conditions:
\begin{itemize}
    \item \textbf{Middle-Group  $M_{P,U}$}: Shares two identity markers with the in-group and contrasts on the other two. In our implementation, $M_{P,U}$ shares the painting preference and university with the in-group (markers $x_P$ and $x_U$ are common) but is from a different session and therefore did not share the group task ($x'_S$, $x'_G$). Formally, \[M_{P,U} = [x_P,\;x_U,\;x'_S,\;x'_G]\,.\]
    \item \textbf{Middle-Group $M_{P,S}$}: Shares the painting preference $x_P$ and is also present in the same session $x_S$, but not from the same university $x_U$ and therefore not in the same group $x_G$, \[M_{P,S} = [x_P,\;x'_U,\;x_S,\;x'_G]\,.\]
    \item \textbf{Middle-Group $M_U$}: Shares only the university affiliation $x_U$ with the in-group and differs on the remaining three, \[M_U = [x'_P,\;x_U,\;x'_S,\;x'_G]\,.\]
    \item \textbf{Middle-Group  $M_P$}: Shares only one identity marker with the in-group and differs on the remaining three. In our implementation, $M_P$ shares the painting preference $x_P$ with the in-group but is from \textbf{another} university and session, with no shared task ($x'_U$, $x'_S$, $x'_G$). Formally, \[M_P = [x_P,\;x'_U,\;x'_S,\;x'_G]\,.\] 
\end{itemize}

These group definitions directly reflect the identity marker framework from section \ref{framework}, here operationalised with four markers. Table \ref{group-characteristics} summarises the identity marker configurations of the in-group, the out-group, and the four middle-groups.

\begin{table}[H]
\centering
\begin{tabular}{c|c|c|c|c|c|c|c}
    &                           & \textbf{In}           & \textbf{M$_{P,U}$}   & \textbf{M$_{P,S}$}  & \textbf{M$_U$$$}    & \textbf{M$_P$}             & \textbf{Out}    \\ \hline
$x_P$ & \textbf{Painting}  & $\surd$               & $\surd$                   &  $\surd$            &                     & $\surd$                                  &                 \\ \hline
$x_U$ & \textbf{University}     & $\surd$               & $\surd$              &                     & $\surd$             &                                     &                 \\ \hline
$x_S$ &\textbf{Session}         & $\surd$               &                      & $\surd$             &                     &                                     &                 \\ \hline
$x_G$ &\textbf{Group Task}      & $\surd$               &                      &                     &                     &                                     &                 \\ \hline
\end{tabular}
\caption{Overview group characteristics. Members of the in-group share all identity markers with each other. The out-group member share none. The middle-groups share some, but not all, markers.}\label{group-characteristics}
\end{table}

It is important to note that some identity markers in our design are \textbf{nested rather than independent}. In particular, sharing the group task marker $x_G$ inherently implies sharing $x_S$ (since one can only complete the task together by being in the same session). However, the converse is not true: two participants might share the same painting preference ($x_P$), same session ($x_S$) and university ($x_U$) without being in the same task group ($x_G$). Thus, in our framework $x_G$ is the most specific marker. 

We model this nested structure deliberately, as such dependencies between identity dimensions often mirror real-world social groupings. For instance, engaging in a joint group task ($x_G$) typically presupposes co-presence in time ($x_S$) and institutional affiliation ($x_U$). Rather than treating identity markers as artificially orthogonal, our design reflects how layered identities naturally co-occur. Notably, the experimental setup still defines groups in clearly distinct and non-overlapping practical categories (e.g., no treatment pair individuals who share $x_G$ but not $x_S$).

Finally, we emphasise that \textbf{marker strength may matter as much as (or more than) the count of shared markers}. While our framework can classify groups by the quantity of overlapping markers, the nature of each marker can differentially impact psychological salience. For example, $x_P$ (painting preference) is a trivial, minimal-group marker, whereas $x_U$ (university affiliation) reflects a meaningful real-world identity. Likewise, $x_G$ (shared group task) creates a direct cooperative bond that might carry more weight than a merely contextual commonality. 

All group labels ($In$, $M_{P,U}$, $M_P$, $M_U$, $M_{P,S}$ $Out$) were \textbf{not explicitly mentioned to participants}; instead, participants saw descriptive phrases for each cup (e.g., “someone at your [same university], from a different session who chose [the same artist]” for a middle-group $M_{P,U}$ member). This ensured that any differential behaviour truly stems from the manipulated identity markers rather than explicit category labels.

\subsection{Altruistic Action}
The term ``altruistic behaviour" describes a wide range of behaviours.
In this study, we focus on a specific kind: \textbf{the act of breaking a small, unenforceable, and unobservable rule for the benefit of another individual.} 
This allows us to investigate altruism under conditions where \textbf{behaviour is observed rather than self-reported} and where the \textbf{act does not impact the individual's reputation}, removing social incentives.
Therefore, there is no personal material gain from the altruistic act, either directly or indirectly, even when directed towards an in-group member.

\subsection{Task}
To operationalise our focus on small rule-breaking for the benefit of others, we adapt the ``Mind Game'' or resource allocation game, previously used to explore moral decision-making and cheating behaviour \citep{greene2009patterns,jiang2013cheating, hruschka2014impartial, purzycki2016moralistic}. Our approach also connects to experimental work on how group identity affects rule-breaking behaviour in economic games. \cite{benistant2019unethical} find that priming group identity increases the prevalence of norm violation in competitive settings, suggesting that group membership can shift the willingness to deviate from prescribed rules.
For a methodological discussion of die-based approaches to measuring dishonesty, see \cite{hermann2025card}.

In step 1, participants are instructed to mentally choose a cup as represented in table \ref{MidCupGroup} (Instruction: ``Step 1: Choose one of the two cups in your mind."). Depending on their current round and treatment, they either see only two cups, which we refer to as in- and out-group, or three cups, including a middle-group.
In step 2, they are instructed to roll a die, either physically at home or through an app or website (Instruction: ``Step 2: Roll the die once.").
In step 3, they are instructed to allocate a coin based on rules that prescribe into which cup the coin should be allocated based on the mentally chosen cup and the outcome of the die roll (Instruction: ``Step 3: Allocate the coin."). For example, for two cups, if the die shows 1, 2 or 3, they should put the coin into the cup they chose in step 1; otherwise, into the other cup. All instructions, including the rules for 3 cups, are outlined in appendix \ref{instr}.
Therefore, participants can always follow or break the rule, as nobody can observe the cup they chose or the outcome of the die roll (if they rolled a die at all).

This process is repeated 30 times in part 1 and 30 times in part 2. In some treatments, the group composition changes in part 2, as will be explained in the next section.

This design satisfies our criteria for studying small, unenforceable, and unobservable rule-breaking. 
The procedure and rules make it impossible to identify a single decision as rule-following or rule-breaking. But because of the task's stochastic nature, we can test the aggregate results against the theoretical, rule-following behaviour. Through this comparison and statistical analysis, we can measure the effects of treatment. Therefore, the task measures biases towards groups through actual behaviour. This is potentially less subject to experimenter demand effect than other tasks which explicitly prompt participants to give allocations to each group in one activity (e.g., \textit{You have 30 coins available, please indicate how many coins you want to give to in-, middle- and out-group."}).

\subsection{Procedure and Treatments}

\paragraph{Session Procedure:}
At the start of the session, participants are provided a link to the experiment environment on oTree via Zoom, where audio, video and general chat are disabled for participants. The default session size was 30 participants. Some sessions had fewer sign-ups, but always had at least 10 participants to allow for the group matching.
As part of the demographics questions, participants indicate their university. 
In the experiment, participants first choose whether they prefer a Klee or a Kandinsky painting. 
Participants are randomly grouped based on their painting preferences and university affiliation and then randomly assigned to one of the treatment conditions. The default group size was 3 participants, but could be increased to 5 if the group matching required it due to the participants' painting preferences.
Each group carries out the group task explained above. 
Following this, instructions for part 1 are provided, and participants individually engage in the task for 30 rounds at their own pace. 
Subsequently, they receive instructions for part 2, engaging in another set of 30 rounds.
After the end of the experiment, participants had the opportunity to sign up for voluntary post-experimental ethnographic interviews where the qualitative answers could not be linked to the quantitative results to allow for total anonymity and transparency. On the one hand, these interviews served as an interdisciplinary research project with anthropology; on the other hand, in the discussion section, we will present a potential explanation for a quantitative finding with insights from the interviews.\footnote{The interviews were conducted by Melina Hühn and supervised by Prof. Dr. Guido Sprenger, Institute of Anthropology, Heidelberg Universtiy.}

\paragraph{Treatments:}
Our study incorporates three distinct cup setups:

\begin{itemize}[itemsep=0pt]
    \item \textbf{In / Out:} This setup involves a cup labelled for the in-group and another labelled for the out-group.
    \item \textbf{In / Middle$_x$ / Out:} This setup includes a third middle cup for group $M_x$,  alongside the in-group and out-group cups.
\end{itemize}

Table \ref{MidCupGroup} shows these cups as they were presented to the participants. The group concepts in-, middle- and out-group were never mentioned but described as outlined above. The labels $\#$, $*$ and $<>$ were used to identify the individual cups. 

\begin{table}[H]
\centering
\begin{tabular}{c||c|c|c}
      \textbf{Group Concept} & \textbf{In} & \textbf{(Middle$_{P,U}$)} & \textbf{Out} \\ \hline \hline
 \raisebox{4\height}{\textbf{Participant View}} & 
 \includegraphics[width=0.1\linewidth]{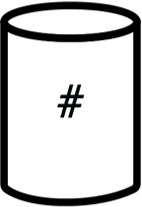} & 
 \includegraphics[width=0.1\linewidth]{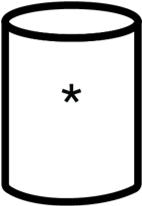} & 
 \includegraphics[width=0.1\linewidth]{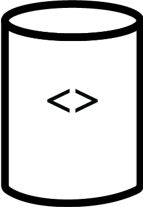} \\ \hline
\textbf{\makecell{Description for \\ Participants} } & \makecell{ \scriptsize Someone from your group \\ \scriptsize in this session at \\ \scriptsize UNI where everyone  \\ \scriptsize chose ARTIST. } & \makecell{\scriptsize Someone at UNI, from \\ \scriptsize a different session who \\ \scriptsize chose ARTIST.}  & \makecell{ \scriptsize Someone not at UNI and \\ \scriptsize not in this session \\ \scriptsize and who chose the \\ \scriptsize other painting.}

\end{tabular}
\caption{Cup association with group concepts}\label{MidCupGroup}
\end{table}

Table \ref{treat} provides the overview of our treatment conditions.

\begin{itemize}[itemsep=0pt]
    \item \textbf{T1 (2-2):}        Is a stable group setting consisting of an in- and out-group.
    \item \textbf{T2 (2-3$_{P,U}$):} Starts with an in- and out-group in part 1, and additionally introduces a middle-group $M_{P,U}$ in part 2.
    \item \textbf{T3 (2-3$_{P,S}$):} Starts with an in- and out-group in part 1, and additionally introduces a middle-group $M_{P,S}$ in part 2.
    \item \textbf{T4 (2-3$_U$):} Also starts with an in- and out-group in part 1, and additionally introduces the more distant middle-group $M_U$ in part 2.
    \item \textbf{T5 (2-3$_P$):} Also starts with an in- and out-group in part 1, and additionally introduces the more distant middle-group $M_P$ in part 2.
    \item \textbf{T6 (3-3$_{P,U}$):} Is a stable group setting consisting of an in-, middle- and out-group. The middle-group $M_{P,U}$ is present here both in parts 1 and 2.
\end{itemize}

\begin{table}[H]
\centering
\begin{tabular}{l||c|c||c}
\textbf{Treatment}       & \textbf{Rounds 1 - 30} & \textbf{Rounds 31 - 60}      & \textbf{N}       \\ \hline \hline
\textbf{T1} (2-2)       & In / Out                & In / Out               & 59      \\ \hline
\textbf{T2} (2-3$_{P,U}$)       &    In / Out                            & In / Middle$_{P,U}$ / Out    & 63                        \\ \hline
\textbf{T3} (2-3$_{P,S}$)       &    In / Out                            & In / Middle$_{P,S}$ / Out    & 63                        \\ \hline
\textbf{T4} (2-3$_U$)   & In / Out                 & In / Middle$_U$ / Out  & 69 \\ \hline
\textbf{T5} (2-3$_P$)   & In / Out                 & In / Middle$_P$ / Out  & 62 \\ \hline
\textbf{T6} (3-3$_{P,U}$)       & In / Middle$_{P,U}$ / Out                & In / Middle$_{P,U}$ / Out  & 60       \\

\end{tabular}
\caption{Treatment overview} \label{treat}
\end{table}

\paragraph{Earnings:} The participation fee for all participants to take part in the lab-in-zoom was 2€. Solving the group task successfully was rewarded with 2€ per person. To highlight the common reward as a group, this was communicated as: ``In the next activity, your group can earn [20 * group-size] coins. The reward will be equally split, so that you individually will earn 20 ECU if your group solves the task correctly. 
Each coin / ECU in the experiment was worth 0.10€. Every participant distributed 60 coins (6€) over the course of the experiment. At the end, every participant randomly receives a ``coin-box" from another participant of their in-, middle- (if applicable) and out-group for each part. Due to a slight imbalance in session size, this did not result in an exact 1:1 matching, which would have meant that all participants would distribute 60 coins and also receive, on average, 60 coins. They received slightly more, which yielded an average payment of 2€+2€+6.23€=10.23€.

\subsection{Operationalisation of Rule Breaking}

To examine rule-breaking behaviour, we record the number of coins placed into a cup by a participant and compare it to the expected number of coins for each part of 30 rounds. 
The expected number of coins is 50\% of the total coins for two cups and 33\% of the total coins for three cups, i.e. 15 and 10 coins over 30 rounds.
 
As our variable of interest, we measure the extent of deviation from expected values using the percentage deviation, $\%\Delta coins = \frac{ActualCoins - ExpectedCoins}{ExpectedCoins}$. 
For example, a $\%\Delta coins$ value of $0.1$ indicates that for every coin that should have been placed in a particular cup, an additional 10\% of coins were put in that cup by that participant.

\section{Results}
In this section, we first confirm that in-group favouritism is consistent across all treatment groups.
We then examine T1 (2-2), which featured only in- and out-groups throughout the experiment.
Next, we analyse treatments T2 and T3, where middle-groups shared two identity markers with the in-group.
This is followed by treatments T4 and T5, where middle-groups shared only one identity marker.
Finally, we turn to T6 (3-3), where a middle-group with two shared identity markers was present in both parts of the experiment.
We conclude with a consolidation of all middle-group results.

\subsection*{In-Group Favouritism is Consistent Over All Treatment Groups}

\begin{figure}[htbp]
    \centering
    \includegraphics[width=0.8\linewidth]{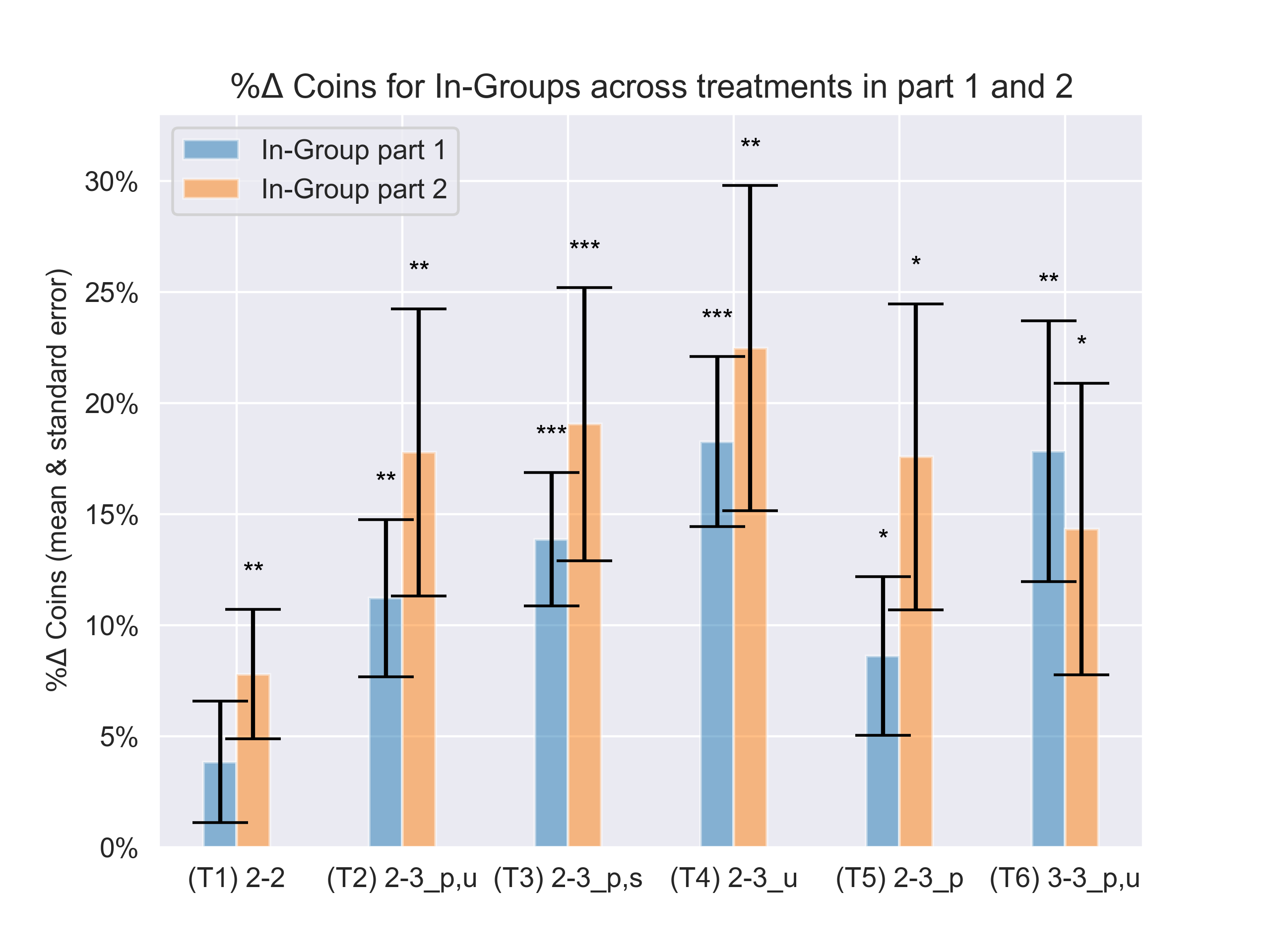}
    \caption{Mean (±SE) of $\%\Delta coins$ allocated to the in-group in each treatment (T1–T6) for Part 1 (blue bars) and Part 2 (orange bars). The theoretical no-favouritism benchmark is at $\%\Delta coins = 0$. Stars indicate significant deviations from 0 (Wilcoxon Signed-Rank tests; $*$ for $p<0.05$, $**$ for $p<0.01$, $***$ for $p<0.001$). In all cases except Part 1 of T1, in-group allocations are significantly above 0, confirming the presence of in-group favouritism.}
    \label{in-group-perct}
\end{figure}

First, we measure in-group favouritism as the $\%\Delta coins$ that go additionally to the in-group.
Looking at figure \ref{in-group-perct}, we see that across all treatments and parts, the mean $\%\Delta coins$ allocated to the in-group is positive. Except for part 1 of Treatment 1, these positive deviations are significantly greater than 0, the theoretical benchmark (Wilcoxon Signed-Rank one-sample tests against 0). 
Here it is important to note, that treatments T1 - T5 are identical in this part 1 and we can therefore pool those observations and yield a highly significant deviation from zero ($p<0.001$ with a Wilcoxon Signed-Rank Test against the $0$ benchmark). 
This tells us that, irrespective of the composition and timeline of the group setting developed in table \ref{imo-line}, we always observe in-group favouritism expressed in rule-breaking for the benefit of the in-group.
Furthermore, the mean in part 2 exceeds that of part 1 for all treatments except for treatment 6 (the ``3-3" treatment), which we will discuss further when we analyse the effect of the middle-group. 

\subsection*{T1 (2-2): Stable In- and Out-Group Over Time}

\begin{figure}[htbp]
\centering
\begin{minipage}{.5\textwidth}
\centering
\includegraphics[width=\linewidth]{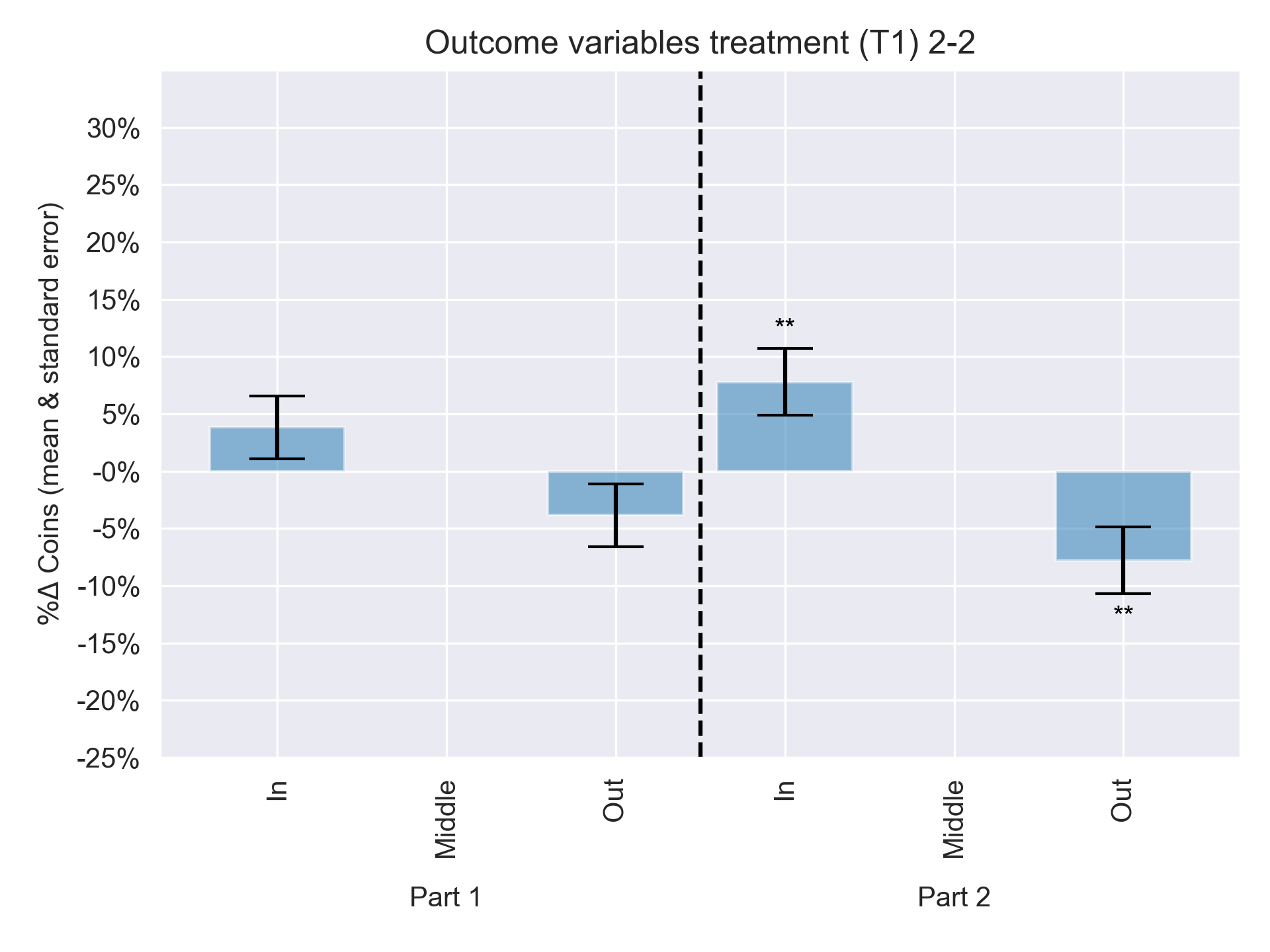}
\end{minipage}%
\begin{minipage}{.5\textwidth}
\centering
\includegraphics[width=\linewidth]{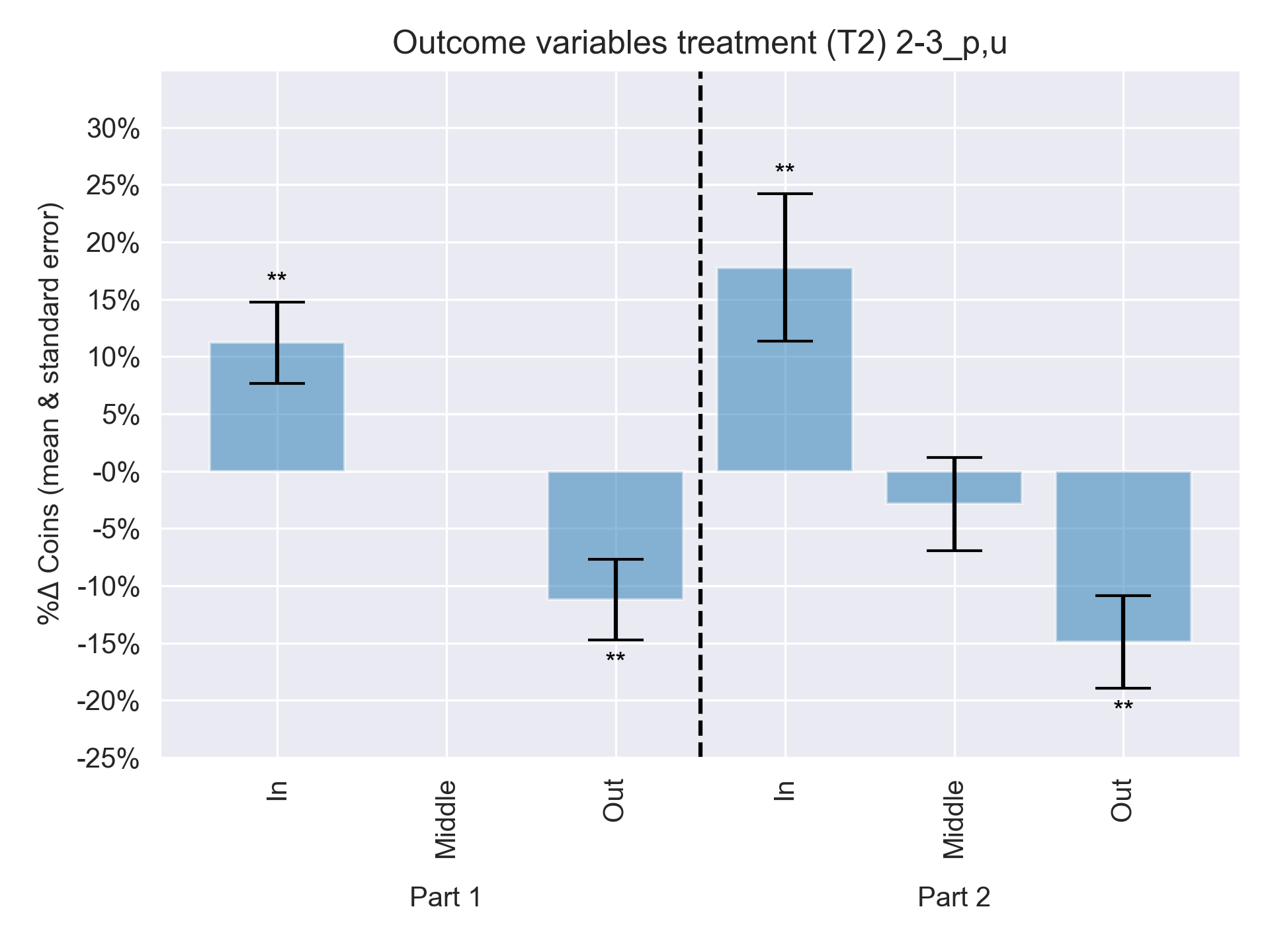}
\end{minipage}

\medskip

\begin{minipage}{.5\textwidth}
\centering
\includegraphics[width=\linewidth]{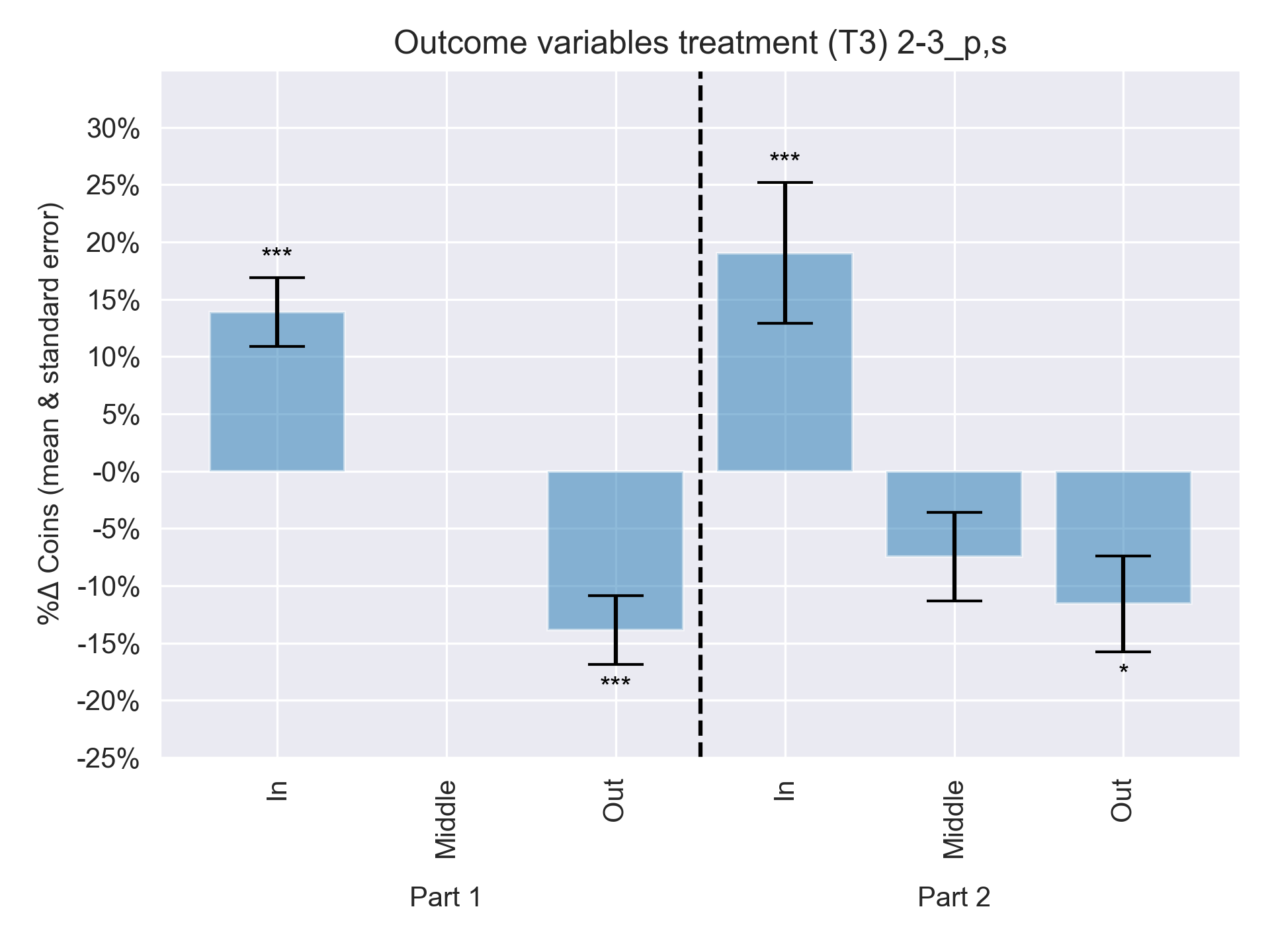}
\end{minipage}%
\begin{minipage}{.5\textwidth}
\centering
\includegraphics[width=\linewidth]{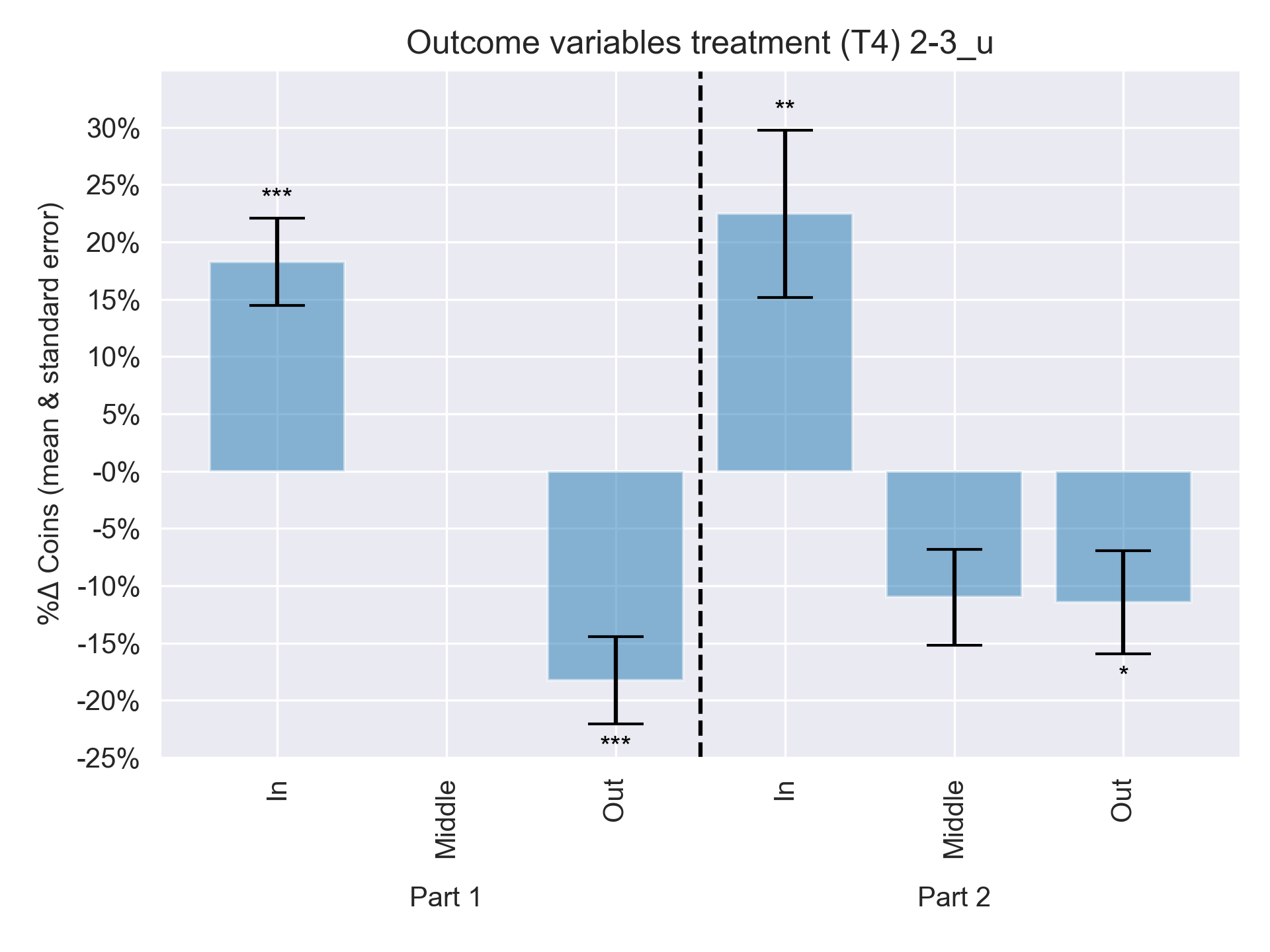}
\end{minipage}

\medskip

\begin{minipage}{.5\textwidth}
\centering
\includegraphics[width=\linewidth]{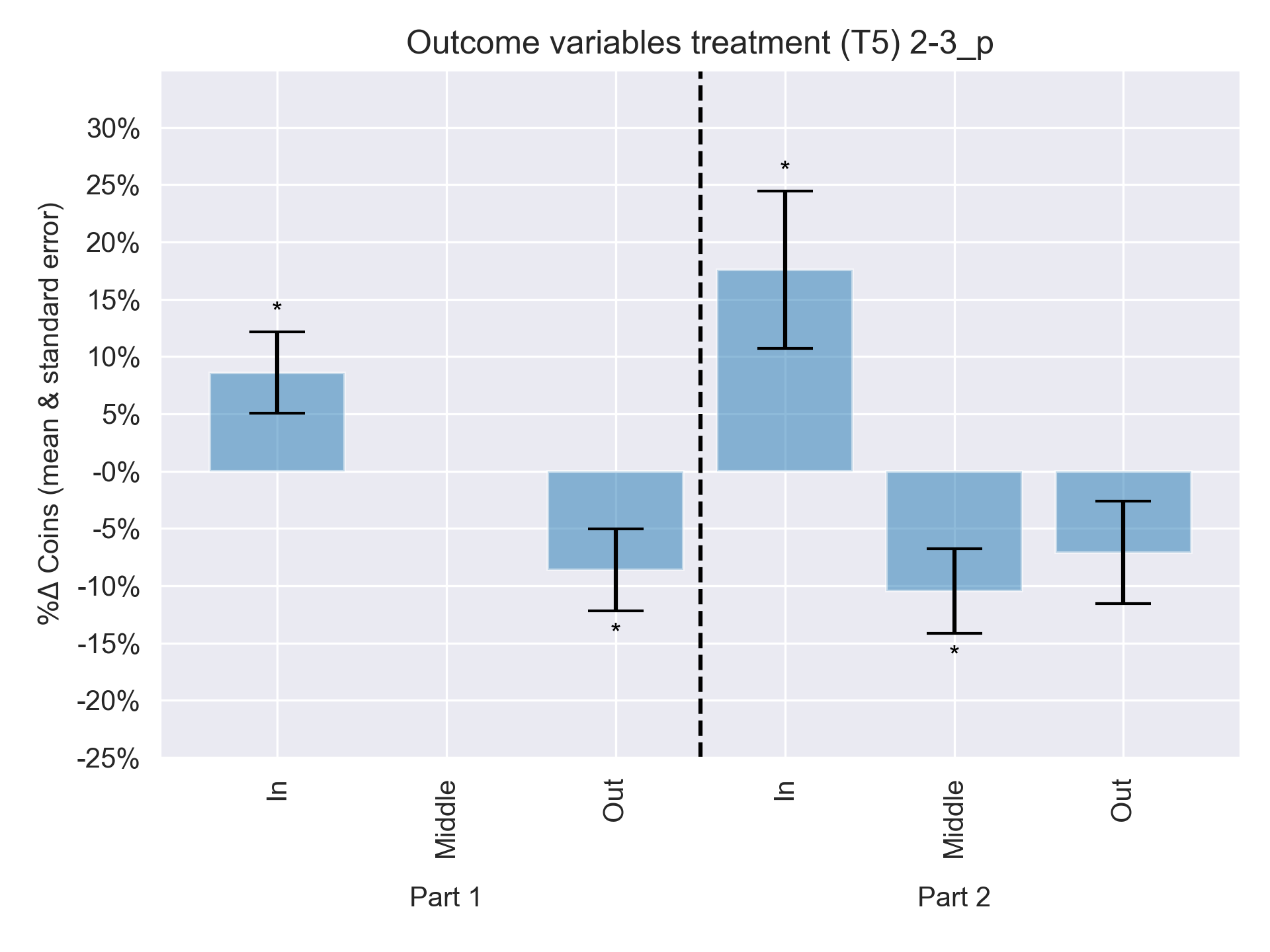}
\end{minipage}%
\begin{minipage}{.5\textwidth}
\centering
\includegraphics[width=\linewidth]{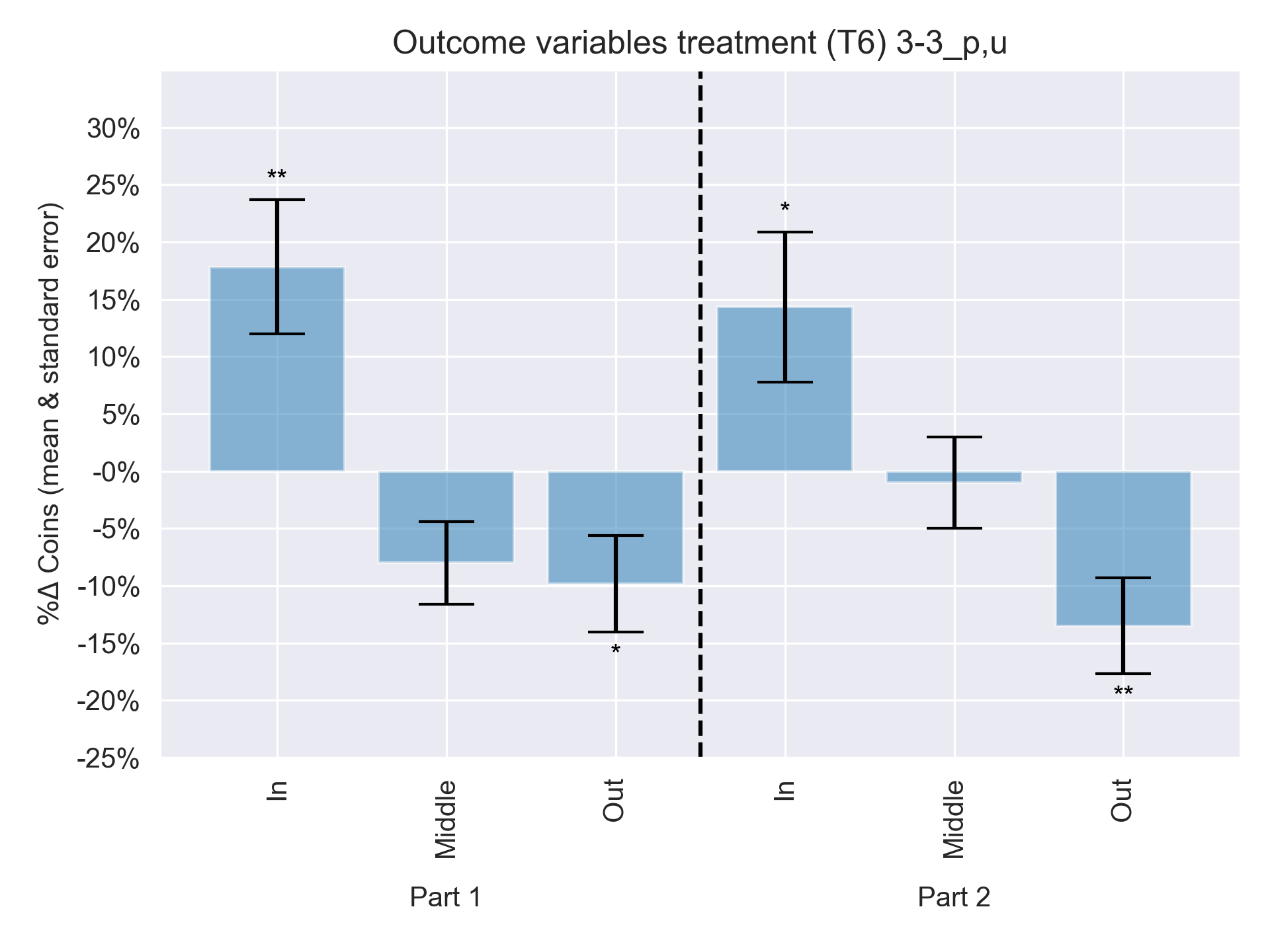}
\end{minipage}
\caption{Mean $\%\Delta coins$ allocated to the in-, middle-, and out-group in each treatment (T1–T6) and experimental part. This represents the $\%$-deviation to the expected value of the rule-following threshold. Error bars show standard errors, and values within each part per treatment sum to zero (zero-sum allocation game). Stars indicate significant deviation from 0 for a given group (Wilcoxon one-sample tests, $*$ for $p<0.05$, $**$ for $p<0.01$, $***$ for $p<0.001$)).
Full results are reported in appendix \ref{app:stats}.}\label{dcoin-all}
\end{figure}

In treatment 1, the group composition stays stable with an in- and out-group in both parts.
Figure \ref{dcoin-all} shows the averages of $\%\Delta coins$ for all treatments and both parts for in-, middle-, and out-group. Within each part per treatment, the means of the groups add up to 0, as the allocation task is a zero-sum game.
For (T1), shown in the top left panel of figure \ref{dcoin-all}, $\%\Delta coins$ for the in-group increases from 3.8\% to 7.8\% in part 2. While in part 2, there is statistically significant rule-breaking in favour of the in-group ($p=0.009$), the difference between part 1 and part 2 is not statistically significant (Wilcoxon signed-rank test for paired data, $p =0.255$).
The value for the out-group is here the inverse, because of the zero-sum nature of the game.

\subsection*{Middle-Groups That Share Two Identity Markers with the In-Group}
Here we analyse T2 (2-3$_{P,U}$) and T3 (2-3$_{P,S}$), whose middle-groups share the painting preference and only differ on the second shared identity marker with the in-group: being in the same university and not in the same session versus not being in the same university, but taking part in the same lab-in-zoom session.

\subsection*{T2 (2-3$_{P,U}$): Introduction of a Middle-Group: A Clear Hierarchy of Groups Emerges}

In treatment 2, the group composition starts with an in- and out-group in part 1. In part 2, a middle group $M_{P,U}$ is introduced that shares two identity markers with the in-group. These identity markers are the arbitrary preferences on Klee vs Kandinsky painting and the lived reality of university affiliation.

Figure \ref{dcoin-all} shows in the top right panel for (T2) that $\%\Delta coins$ for the in-group increases from 11.2\% to 17.8\% in part 2 (not statistically different from each other with Wilcoxon Signed-Rank for paired data, $p=0.420$). 
For the out-group it decreases from -11.2\% to -14.9\%. This is not the inverse of part 1 because of the presence of the middle-group with a mean of -2.9\% in part 2, which is however not significantly different from 0 (Wilcoxon Signed-Rank test against 0, $p = 0.651$). 

In Part 2 of T2, a clear hierarchy of groups emerges that can be statistically confirmed. The in-group is treated significantly more favourably than the middle-group (Mann-Whitney $U = 2420$, $p = 0.016$. And $p = 0.016<0.025$, threshold for significance with Holm correction), and the middle-group is treated significantly more favourably than the out-group ($U = 2342$, $p = 0.040<0.05$, the threshold for significance with Holm correction). 

This shows that when a middle-group that shares sufficient identity markers is introduced, a clear hierarchy of groups emerges. The in-group continues to profit, where the out-group bears the loss, while the middle-group is treated according to the rule (no significant difference against the zero-threshold). As we will discuss below, a question for further research is what a ``sufficient" sharing of identity markers is.

\subsection*{T3 (2-3$_{P,S}$): Not All Markers are the Same - The Middle Group Starts to Lose}

The middle-group in treatment T3 also shares the painting preference with the in-group, is also placed in the same lab-in-zoom session, but is from a different university. Therefore, this middle-group has the same \textit{amount} of shared identity markers, but switches the second shared marker from the same university affiliation to presence in the same session, which is conducted across universities. 

The middle left panel shows that also here in part 2 the in-group is statistically significantly treated favourably, and the out-group is statistically significantly treated hostile. The middle-group is not statistically significantly treated differently from the zero benchmark, yet the average of -7.5\% is visibly lower than the average of -2.9\% in T2. However, unlike in T2, we cannot confirm that this middle-group is treated significantly better than the out-group (Mann-Whitney $U = 2093$, $p = 0.298$). Furthermore, a direct comparison between the middle-groups in T2 and T3 shows no significant difference ($p = 0.408$). Thus, while the descriptive pattern suggests M$_{P,S}$ occupies a middle position between in- and out-group, this cannot be statistically confirmed with the present sample, yet we can statistically confirm that it is not treated differently from the zero-threshold while in- and out-group are.

\subsection*{Middle-Groups that share one identity marker with the In-Group: Clear ``Us (In-Group) versus Them (All Other Groups)"}
Here we analyse T4 (2-3$_{U}$) and T5 (2-3$_{P}$), whose middle-groups each only have one identity marker in common with the in-group: having chosen the same painting, but not in the same university and session, versus being in the same university and not in the same session and not the same painting.

\subsection*{T4 (2-3$_U$): Introduction of Another More Distant Middle-Group - Middle-Group Treated Similar to an Out-Group}

In treatment 4, the group composition also starts with an in- and out-group in part 1. 
Then, in part 2, a middle-group that shares only one identity marker is introduced. The shared identity marker is the university affiliation, while they differ in the other markers, i.e., painting preference and session.

While the behaviour towards the middle-group is not statistically significant ($p=0.060$), the average ($-11.0\%$) is visibly very close to the average of the out-group (-11.4\%), and also visibly close to the average of the middle-group in T5 ($-10.5\%$).

This shows that the markers of \textit{painting} and \textit{university} each individually lead to a behaviour towards the middle-group similar to the behaviour towards the out-group and only jointly lead to a behaviour that treats the middle-group according to the given rule.   

\subsection*{T5 (2-3$_P$): Introduction of a More Distant Middle-Group - Same Behavioural Pattern}

In treatment 5, the group composition also starts with an in- and out-group in part 1. 
Then, in part 2, a middle-group that shares only one identity marker is introduced. The shared identity marker is the shared painting preference, while they differ in the other markers, i.e., university affiliation and session.

Similar to treatment 2, the bottom left panel in figure \ref{dcoin-all} shows an increase for the in-group from 8.6\% to 17.6\%, yet not statistically different in a Wilcoxon Signed-Rank test for paired data ($p=0.104$). 
Yet the introduced middle-group receives a $\%\Delta coins$ of -10.5\%, which is significantly different from 0 (Wilcoxon Signed-Rank test against 0, $p = 0.013$), in contrast to treatment 2, where the middle-group showed no significant deviation. At the same time, the value for the out-group slightly increases from -8.6\% to -7.1\%.

In both T4 and T5, the middle-group is not treated significantly differently from the out-group (T4: $p = 0.727$; T5: $p = 0.080$). This suggests that both the more distant middle-group and the out-group are perceived as out-groups. Behaviour towards these groups is not differentiated based on their relative proximity. It can be summarised as `us (in-group) versus them (all other groups)' behaviour.

This suggests a threshold exists: groups are perceived as more neutral, as in treatment 2, when they share enough identity markers. Conversely, in cases like treatment 4 and 5, `insufficient' shared identity markers result in all non-in-groups being perceived as out-groups.

\subsection*{T6 (3-3$_{P,U}$): Stable, In-, Middle- and Out-Group}

In treatment 6, the group composition stays stable over time, including a `closer' middle-group $M_{P,U}$ that shares two identity markers of painting preferences and university affiliation.

The bottom right panel in figure \ref{dcoin-all} shows that in part 1, the in-group receives 17.8\% in $\%\Delta coins$ (significantly above 0, Wilcoxon $p=0.007$, while both the middle- ($-8.0\%$, Wilcoxon $p=0.061$) and out-group ($-9.8\%$, Wilcoxon $p=0.047$) are both below zero. This again can be described as an `us (in-group) versus them (all other groups)' behaviour. This suggests that in the beginning of any \textbf{new} environment (e.g., the start of the experiment), participants do not differentiate more than a simple in- and out-group effect. 

Yet in part 2, they clearly do differentiate more, and do not treat the middle-group as an out-group. Here, the in-group receives 14.3\% (Wilcoxon $p = 0.017$ from 0, the middle-group receives -1.0\% which is first of all not significant deviation from 0 ($p= 0.344$), but it is significantly different from the result of the middle-group in part 1 with a Wilcoxon Signed-Rank test for paired data ($p=0.031$). The out-group receives -13.5\% ($p = 0.007$ against the 0 benchmark). This mirrors the results of treatment 2 in part 2 and shows a clear hierarchy of groups in terms of relative proximity exists \textbf{after} experimental participants have had repeated exposure to the experimental setting. In Part 2 of T6, the middle-group is treated significantly better than the out-group (Mann-Whitney U, $p = 0.005$). While the pattern of means (In: +14.3\%, Middle: -1.0\%, Out: -13.5\%) is consistent with a hierarchy, the direct comparison between in-group and middle-group does not reach significance (p = 0.152). Thus, the hierarchy is partially but not fully confirmed in T6.

\subsection*{Consolidating the Results of All Middle-Groups}

\begin{figure}[htbp]
    \centering
    \includegraphics[width=0.9\linewidth]{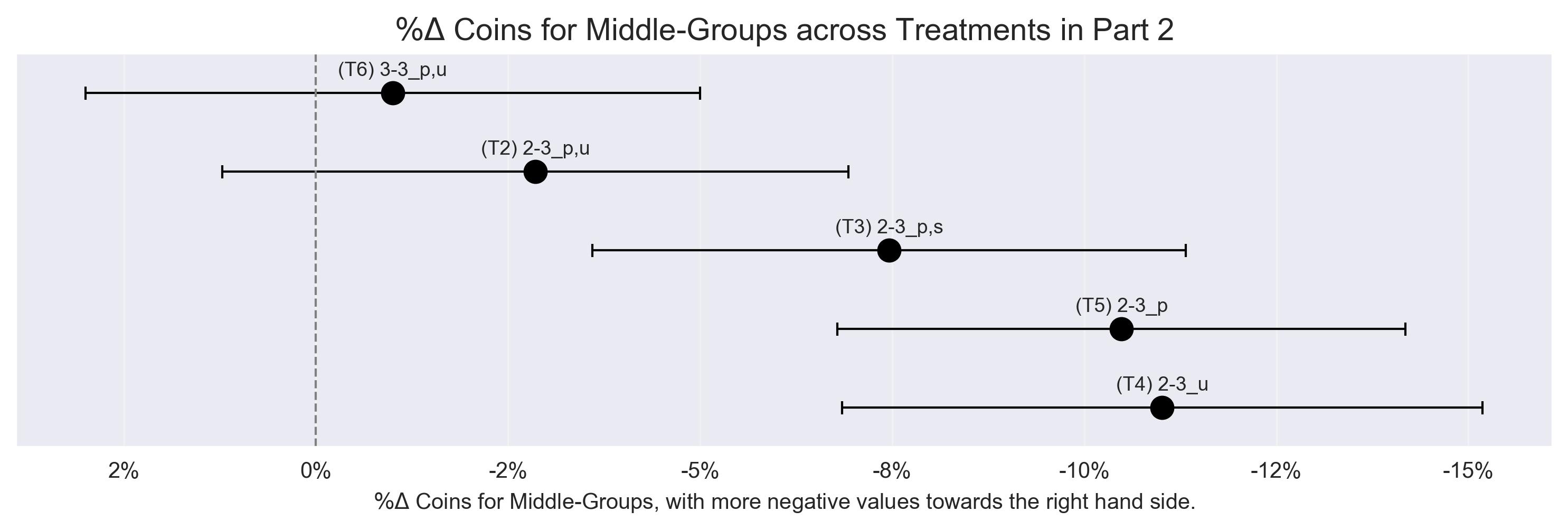}
    \caption{Mean percentage of coins allocated to middle-groups across treatments in Part 2, ordered by mean. Higher values (left) indicate treatment effects closer to the in-group; lower values (right) suggest distance from the in-group. Error bars represent standard errors.}
    \label{middle-part2}
\end{figure}

Figure \ref{middle-part2} consolidates all results for the middle-groups in part 2, from figure \ref{dcoin-all}. It shows the mean and standard error, ordered by mean, where higher means (i.e., closer to zero and a rule-following treatment) appear on the left and more negative means appear toward the right. With this overview, we can assess the empirical placement of each group, as introduced in table \ref{imo-line}. A group shown further to the left is placed closer to the in-group, while a group further to the right is placed closer to the out-group.

We observe that treatment 6, where group $M_{P,U}$ has been present from the beginning, is the closest to the in-group and rule-following benchmark. Following this is treatment 2, where $M_{P,U}$ was introduced only in the second part. It is the only middle-group whose error bar overlaps the zero benchmark. Further to the right is $M_{P,S}$, a middle-group that also shares two markers with the in-group, but where the shared marker refers to a different contextual cue: real-world affiliation (same university) rather than co-presence in the experimental session. Finally, the groups on the far right of the plot, with averages of around -11\%, share only one marker with the in-group and are clearly placed closer to the out-group. A Mann-Whitney U test between T2 and T3 for the allocation to the middle-group in part 2 shows no significant difference. Additionally, a Mann-Whitney U test between T4 and T5 shows no significant difference.

Table \ref{tab:delta-second-middle-treatments} presents an OLS regression on $\%\Delta coins$ middle-group part 2 in treatments 2-5. Here, the identity markers are explanatory variables that can also occur in interaction. We see that only the interaction \textit{Painting $\times$ University} has a significant result, which represents treatment T2 with $M_{P,U}$.

Table \ref{tab:delta-second-middle-treatments} presents an OLS regression on $\%\Delta coins$ for the middle-group in part 2 across treatments 2–5. The model is estimated without intercept, allowing each coefficient to represent the mean allocation to that treatment's middle-group relative to the expected-value benchmark of zero. Only the middle-group in (T2) 2-3$_{P,U}$, which shares both painting preference and university affiliation with the in-group, receives allocations not significantly different from the benchmark (p = 0.477). In contrast, the middle-groups sharing only one marker ((T4) 2-3$_{U}$  and (T5) 2-3$_{P}$) are treated significantly below the benchmark (p = 0.004 and p = 0.010), consistent with out-group-like treatment.

\begin{table}[ht!]
\centering
\begin{tabular}{@{}lcccccc@{}}
\toprule
                            & Coef.   & Std. Err. & t       & $P>|t|$   & [0.025   & 0.975] \\
\midrule
(T2) 2-3$_{P,U}$            & -0.0286 & 0.040     & -0.713  & 0.477   & -0.108   & 0.050 \\
(T3) 2-3$_{P,S}$            & -0.0746 & 0.040     & -1.861  & 0.064   & -0.154   & 0.004  \\
(T4) 2-3$_{U}$              & -0.1101 & 0.038     & -2.875  & 0.004$^{**}$   & -0.186   & -0.035  \\
(T5) 2-3$_{P}$              & -0.1048 & 0.040     & -2.594  & 0.010$^{*}$   & -0.184   & -0.025  \\
\midrule
\multicolumn{7}{@{}l}{\textbf{Model Information:}}\\
\multicolumn{2}{l}{R-squared}       & \multicolumn{1}{r}{0.010} & \multicolumn{4}{r}{}\\
\multicolumn{2}{l}{Adj. R-squared}  & \multicolumn{1}{r}{-0.001} & \multicolumn{2}{l}{No. Observations}& \multicolumn{1}{r}{257} & \multicolumn{1}{r}{}\\
\multicolumn{2}{l}{F-statistic}     & \multicolumn{1}{r}{0.8868} & \multicolumn{2}{l}{Df Residuals}    & \multicolumn{1}{r}{253} & \multicolumn{1}{r}{}\\
\multicolumn{2}{l}{Prob (F-statistic)} & \multicolumn{1}{r}{0.448} & \multicolumn{2}{l}{Df Model}        & \multicolumn{1}{r}{3} & \multicolumn{1}{r}{}\\
\multicolumn{2}{l}{Log-Likelihood}  & \multicolumn{1}{r}{-68.381} & \multicolumn{2}{l}{Covariance Type} & \multicolumn{1}{r}{Nonrobust} & \multicolumn{1}{r}{}\\
\bottomrule
\end{tabular}
\caption{OLS Regression Results on \textbf{$\%\Delta coins$ middle-group part 2}. Each coefficient represents the mean $\%\Delta coins$ for the respective treatment's middle-group. ($*$ for $p<0.05$, $**$ for $p<0.01$, $***$ for $p<0.001$)}\label{tab:delta-second-middle-treatments}
\end{table}

We also estimate a specification including demographic controls (Table \ref{tab:delta-second-middle-controls}). Given the sample size relative to the number of control categories, statistical power and interpretability is reduced.

\section{Discussion \& Conclusion}

In this study, we investigated altruistic rule-breaking across various group settings.  
We designed and controlled these group settings in a multi-lab, lab-in-zoom study, using identity markers derived from real-world associations, the minimal group paradigm and randomness within the lab. 
This allowed us to create an in-group and an anonymous out-group of distant strangers and middle-groups that share some identity markers with the in-group yet are clearly distinct from it.
The altruistic rule-breaking in a die roll game in complete privacy is measured through aggregate analysis.

Our findings confirm the general presence of in-group favouritism. 
Furthermore, we can report a clear pro in-group behaviour when a middle- and out-group are present in the \textit{initial part} of the experiment. Yet this ``us versus them" behaviour at the start evolves into a behaviour reflecting a hierarchy of groups in the second part of the experiment.  
Here, we observe a pro in-group and neutral towards middle-group behaviour compared to a rule following statistical benchmark. Accordingly, the out-group carries the loss in our zero-sum game setting.

This rule-following and neutral behaviour towards a middle-group is also confirmed when such a group is only introduced in the second part of the study.
At the same time, participants show sensitive behaviour to the composition of identity markers of such a middle-group. When a middle-group is introduced that shares the same number of identity markers with the in-group, but is lacking the identity marker based on lived experience (university affiliation), then it is treated less favourably, but still better than an out-group. When the middle-group shares only one marker with the in-group, then it is treated like an out-group, regardless of which of the markers is shared. This creates a clear ``us versus them" behaviour again.

Notably, sharing the university affiliation, arguably the most salient identity marker rooted in lived experience, is not sufficient on its own to differentiate a middle-group from an out-group. Only when combined with an additional shared marker does neutral treatment emerge. This suggests that a single shared identity, however meaningful, may not cross a threshold of perceived similarity required to attenuate out-group categorisation. Participants may require multiple convergent signals before adjusting their behaviour away from a binary in-group/out-group distinction. This is particularly striking given that university affiliation has been shown to generate strong group identity effects in other economic decision-making contexts \citep{chen2022competition}. That this marker alone does not suffice to shift behaviour away from the binary default in our setting reinforces the threshold interpretation.

The finding that differentiated treatment of the middle-group is confined to Part 2 is consistent with cognitive resource accounts of social categorisation. \cite{fiske1990continuum} argue that differentiated impression formation requires both motivation and available cognitive capacity; when capacity is constrained, perceivers fall back on simpler categorical distinctions. \cite{spears1999effect} show more specifically that maintaining complex multi-group representations is itself resource-consuming, and that social categorisation can decrease under cognitive load. In Part 1 of our experiment, participants simultaneously learn a novel task interface, process allocation rules, and encode an unfamiliar three-group structure — conditions that plausibly exhaust the cognitive resources needed for fine-grained group differentiation. By Part 2, the procedural aspects of the task are familiar, freeing capacity for processing the group structure in greater detail. This interpretation is further supported by work showing that when social markers are complex rather than binary, individuals tend to default to idiosyncratic binary decision rules \citep{hertz2025increased}. In our setting, the binary simplification in Part 1 (treating anyone who is not in-group as out-group) may reflect exactly this kind of default. That the shift occurs at the part boundary rather than gradually across rounds (including in the treatment, where three cups are present throughout) points to the re-reading of instructions between parts as a discrete recategorisation prompt rather than to incremental learning within a part. The role of repeated exposure also connects to experimental evidence on inter-group contact. \cite{lenz2022effect} find that cooperative interaction with out-group members reduces taste-based discrimination. In our setting, the repeated encounter with the group constellation between parts may serve an analogous function, prompting participants to process group distinctions with greater care.

These findings have multiple implications for future experimental research on intergroup behaviour.
This study suggests that in one-shot interactions or short experiments with complex group settings, participants show strong in-group favouritism. Yet, after re-encountering the experimental instructions and group constellation in a subsequent phase, participants show more nuanced behaviour.
Furthermore, we present a group setting going beyond the classical in- and out-group dichotomy using several identity markers.
Employing multiple identity markers to go beyond that dichotomy could significantly benefit research on intergroup relations.

Here, we see a vast potential for further research. In the present study, we see a middle-group, $M_{P,U}$, treated neutrally, and several middle-groups that are treated similarly to an out-group. This raises questions on the composition of those identity markers, $x$. How important is each single, isolated identity marker? Which role do markers from lived experience play? How are their conjoint effects? Is there a tipping point in markers where a `neutral' behaviour turns `hostile', or would it be a gradual transition?  Answering such questions builds directly on the framework of the present study. In our study it needs two markers, one of which is the association with a salient part of life (university) from outside the lab. It can inform inter group research and important current social issues around discrimination and polarisation, especially in contexts of intersectionality. 

In this study, we transition from a group dichotomy to a linear ordering of groups based on identity markers.
Further research should increase this complexity to settings where a marker $x$ can take more than two values and interact with other markers. Additionally, in the present study, the markers of in- and out-group remained constant. Future research should analyse the dynamics of multiple groups and changing markers $x$ within and across groups, potentially with (partial) uncertainty of markers.

Naturally, this present study carries limitations.
In the task, we implement a comparably high mental cost of rule-breaking by providing a very strict set of rules. 
How the results change, especially towards a middle-group, when the mental costs of rule breaking change (e.g., even lower) is unclear. 
We also do not know whether experimental demand effects might have directed behaviour in a particular direction.
In our study, we have one treatment with a middle-group present in part 1 and 2, and observe a significant change. This leads to important questions about how effects in experiments are influenced by the familiarity and repeated exposure to instructions and a task in the domain of social preferences. Future studies should investigate specifically this question, how the methodological familiarity with a task changes the measurement of social preferences in experiments. 
In our study, participants take part in the privacy of their homes. Would it make a difference to run the study in a lab or in a public place?
In our setting, the number of coins to be allocated is fixed and a zero-sum game, so the results within each experimental part and treatment condition complement each other, especially in settings with two groups. 
Since we cannot \textit{know} for which exact coin allocation a participant was breaking the rule, we must rely on an aggregate analysis level. While it is possible to detect outliers whose behaviour is very unlikely to have been rule-following, any individual behaviour closer to a rule-following behaviour can never be confidently identified as rule-following or breaking.
On the one hand, this true privacy is a huge benefit of the method to elicit true behaviour, but it brings limitations to the analysis. 
What we know from voluntary, ethnographic interviews conducted after the experiment is that some participants stated that they were fully rule following when the die rolls prescribed to give to the in-group, but when it should go to the out-group multiple times in a row, they would stop that ``lucky streak" at some point by giving to the in-group instead.

Lastly, the sample size of approximately 60 participants per treatment was determined based on feasibility constraints and prior studies using similar paradigms. While sufficient to detect within-treatment deviations from the benchmark, this sample size limits power to detect between-group differences, especially given the stochastic nature of the task. 
Consequently, several theoretically motivated comparisons did not reach significance, and confidence intervals in Figures \ref{dcoin-all} and \ref{middle-part2} overlap substantially for some comparisons.
The complete hierarchy is statistically demonstrated in the treatment with middle-group $M_{P,U}$. We view these as meaningful results that survive a noisy measurement context, while acknowledging that effect sizes and boundary conditions require replication with larger samples.

In conclusion, our study robustly demonstrates that participants in an experimental setting perceive beyond the simple dichotomy of in- and out-groups. They exhibit more nuanced behaviours towards complex group settings. Introducing a middle-group can lead to a clear hierarchy of altruistic rule-breaking where a middle-group is treated neutrally. This finding challenges the traditional in-/out-group dichotomy and suggests richer group settings and dynamics. Further research should follow up on this group setting paradigm to better understand the situations we observe in the world around us, to eventually inform policies tackling polarisation and discrimination.





\bibliographystyle{ecca}
\bibliography{references}
\vspace{0.25cm}

\appendix
\renewcommand{\thetable}{A\arabic{table}}
\renewcommand{\thefigure}{A\arabic{figure}}
\setcounter{table}{0}
\setcounter{figure}{0}

\section{Instructions}\label{instr}
The following instructions are shown for a participant from Heidelberg University who chose the painting by Paul Klee and was part of (T2) 2-3$_{P,U}$.

\subsection*{Start}

\begin{enumerate}
    \item \textbf{Informed Consent} 
    \item \textbf{Start Study}
    \item \textbf{Painting Preferences} ``Which painting do you prefer?" - Choice between Klee, A Dry-Cool Garden (1921) and Kandinsky, Landscape with Red Splashes I	(1913) - Painting order randomised between participants
    
\end{enumerate}

\subsection*{Group Task}

\begin{enumerate}
    \item \textbf{Result Painting and preparation group task:} ``The painting you chose was painted by Paul Klee. - You are now in a group with 3 participants in total. Everybody in this group chose the same painting as you. Everybody in this group is currently present in this ongoing session at Heidelberg University."
    \item \textbf{Instructions Group Task:} ``In the next activity your group can earn 60 [20 * group size] coins. The reward will be equally split so that you individually will earn 20 coins if your group solves the task correctly. Remember, after the study the coins will be converted to Euro, where 1 coin equals 0.1€. Everyone in your group chose the painting by Paul Klee. In the next activity your task is to identify paintings that were drawn by Paul Klee as well. You can identify them by their broad similarity in style compared to the painting you chose. Please take a moment to look at the style of the painting." [Selected painting is shown.]
    \item  \textbf{Group Task:} ``Which painting was drawn by Paul Klee?" with button selection for the painting. Selected paintings are shown in table \ref{paintings}.
    \item \textbf{Result:} If successful (which all groups were): ``Congratulations! - Your group has solved this task successfully and earns 60 coins! Your personal payoff out of the group pot is 20 coins."
\end{enumerate}

\begin{table}[h!]
\centering
\begin{tabular}{c l l }
\hline
\textbf{Round} & \textbf{Klee} & \textbf{Kandinsky} \\
\hline
1 & Irma Rossa the Animal Tamer (1918) & St George III (1911) \\
2 & The Lamb (1920) & Landscape (1914) \\
3 & Senecio (Head of a Man) (1922) & Improvisation 26 (1912) \\
4 & Persian Nightingales (1917) & Improvisation 9 (1910) \\
5 & Angelus Novus (1920) & Improvisation 209 (1917) \\
\hline
\end{tabular}
\caption{Paintings by Klee and Kandinsky across five rounds}\label{paintings}
\end{table}

 \subsection*{Instructions and Task Part 1 (2 Cups)}
\begin{figure}[h!]
\centering

\begin{subfigure}{0.48\textwidth}
    \includegraphics[width=\linewidth]{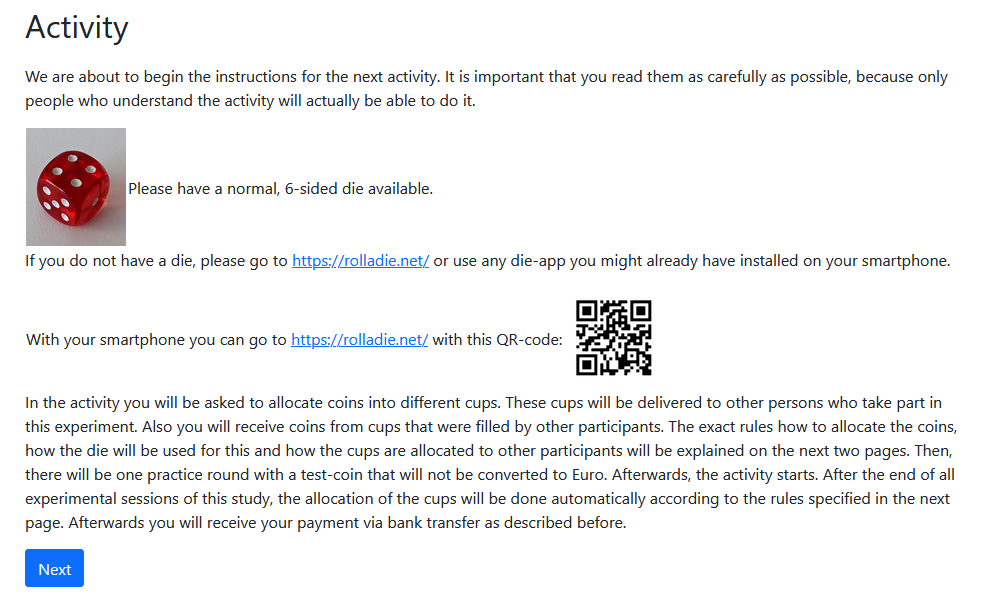}
    \caption{General Instructions}
\end{subfigure}
\begin{subfigure}{0.48\textwidth}
    \includegraphics[width=\linewidth]{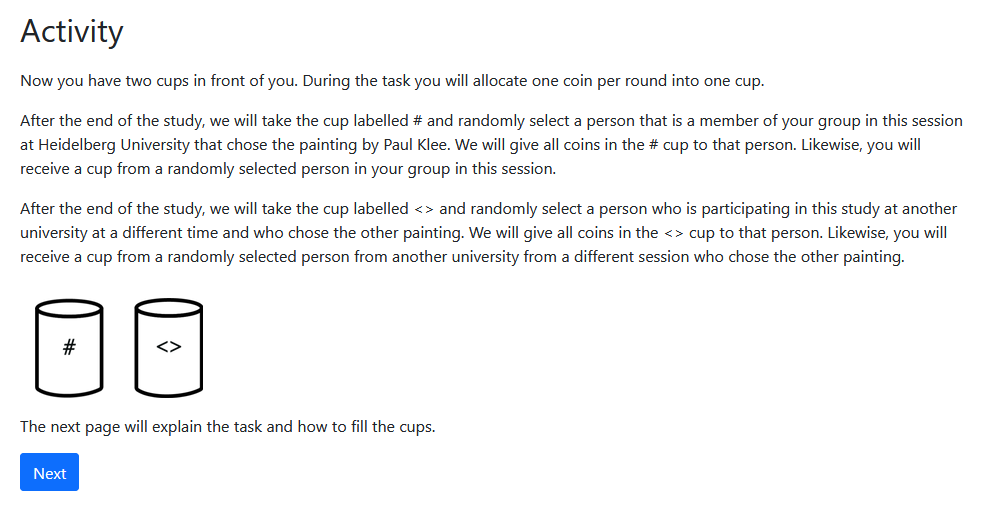}
    \caption{Treatment Instructions (2 Cups) (i)}
\end{subfigure}

\begin{subfigure}{0.48\textwidth}
    \includegraphics[width=\linewidth]{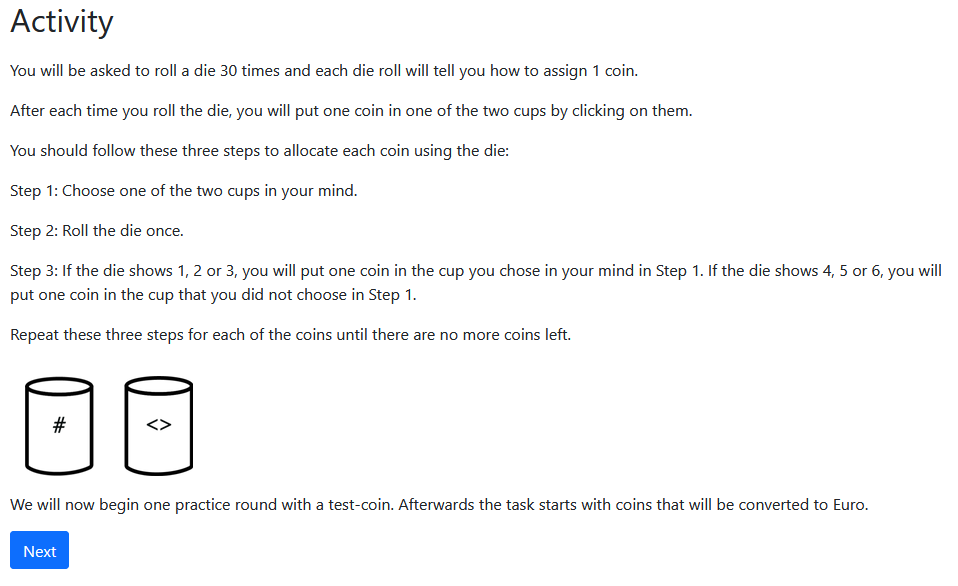}
    \caption{Treatment Instructions (2 Cups) (ii)}
\end{subfigure}
\begin{subfigure}{0.48\textwidth}
    \includegraphics[width=\linewidth]{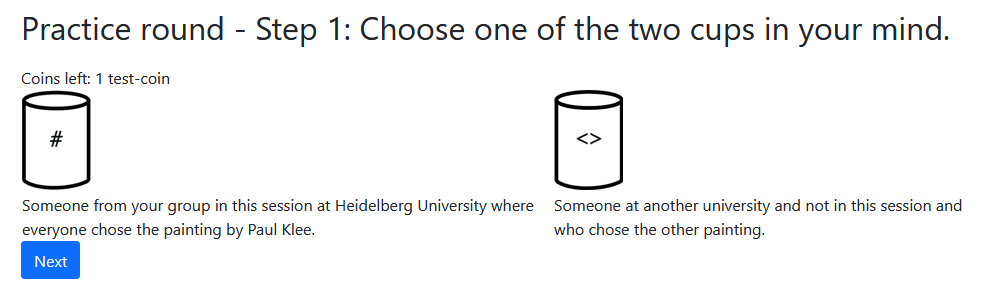}
    \caption{Practice Round (2 Cups) Step 1}
\end{subfigure}

\begin{subfigure}{0.48\textwidth}
    \includegraphics[width=\linewidth]{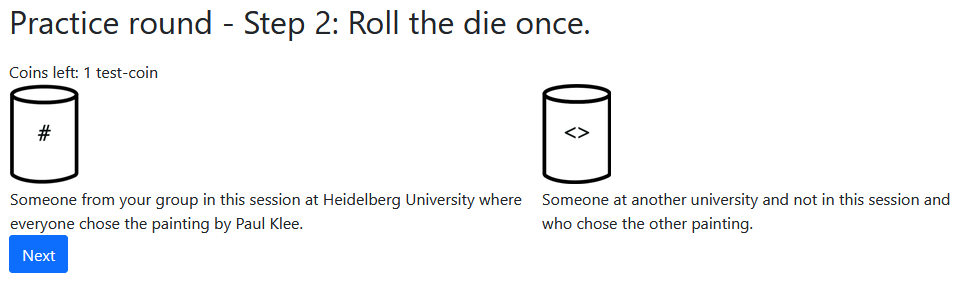}
    \caption{Practice Round (2 Cups) Step 2}
\end{subfigure}
\begin{subfigure}{0.48\textwidth}
    \includegraphics[width=\linewidth]{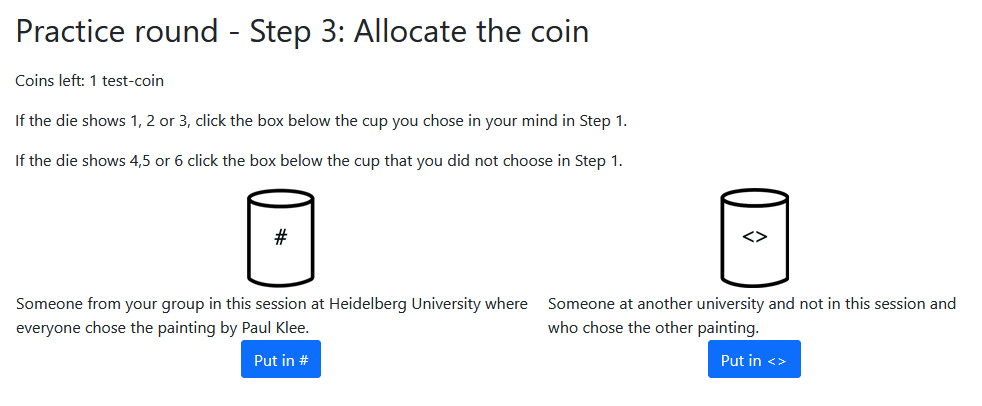}
    \caption{Practice Round (2 Cups) Step 3}
\end{subfigure}

\begin{subfigure}{0.48\textwidth}
    \includegraphics[width=\linewidth]{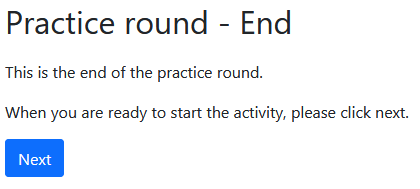}
    \caption{Practice Round End}
\end{subfigure}
\begin{subfigure}{0.48\textwidth}
    \includegraphics[width=\linewidth]{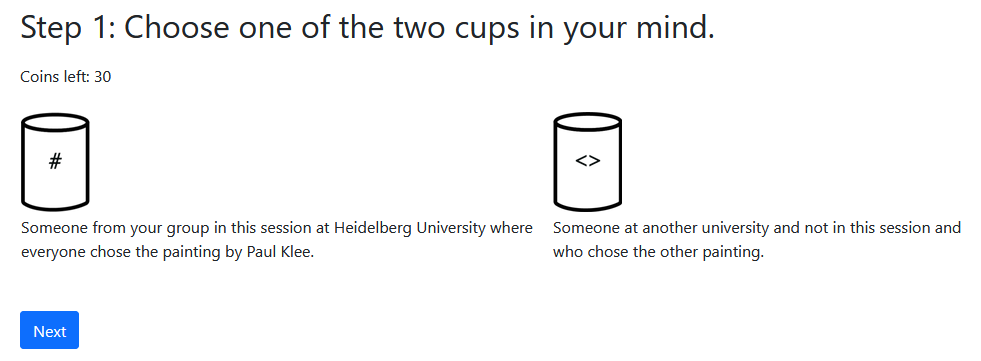}
    \caption{Round 1 (2 Cups) - Step 1}
\end{subfigure}

\begin{subfigure}{0.48\textwidth}
    \includegraphics[width=\linewidth]{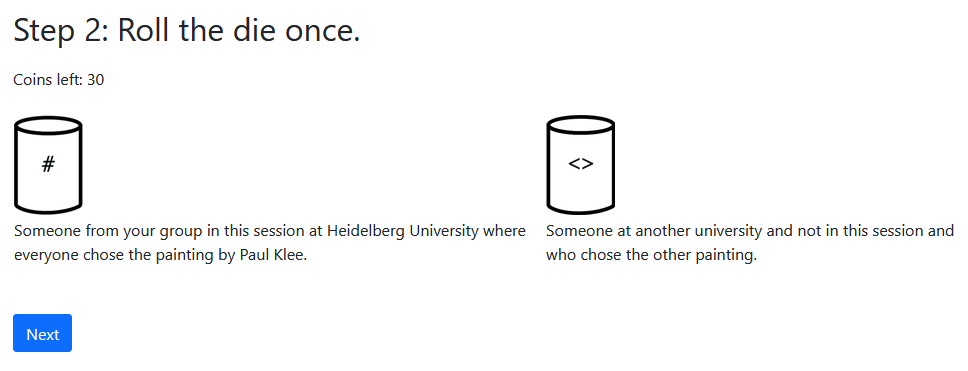}
    \caption{Round 1 (2 Cups) - Step 2}
\end{subfigure}
\begin{subfigure}{0.48\textwidth}
    \includegraphics[width=\linewidth]{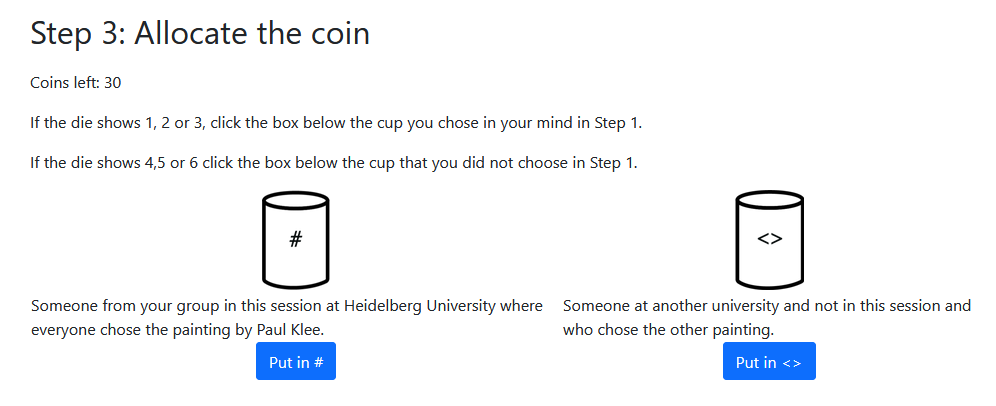}
    \caption{Round 1 (2 Cups) - Step 3}
\end{subfigure}

\caption{Instructions and Task Part 1 (2 Cups)}\label{instr-part1}
\end{figure}

 The instructions are shown in figure \ref{instr-part1}.

 \subsection*{Instructions and Task Part 2 (3 Cups)}
\begin{figure}[h!]
\centering

\begin{minipage}{0.48\textwidth}
    \includegraphics[width=\linewidth]{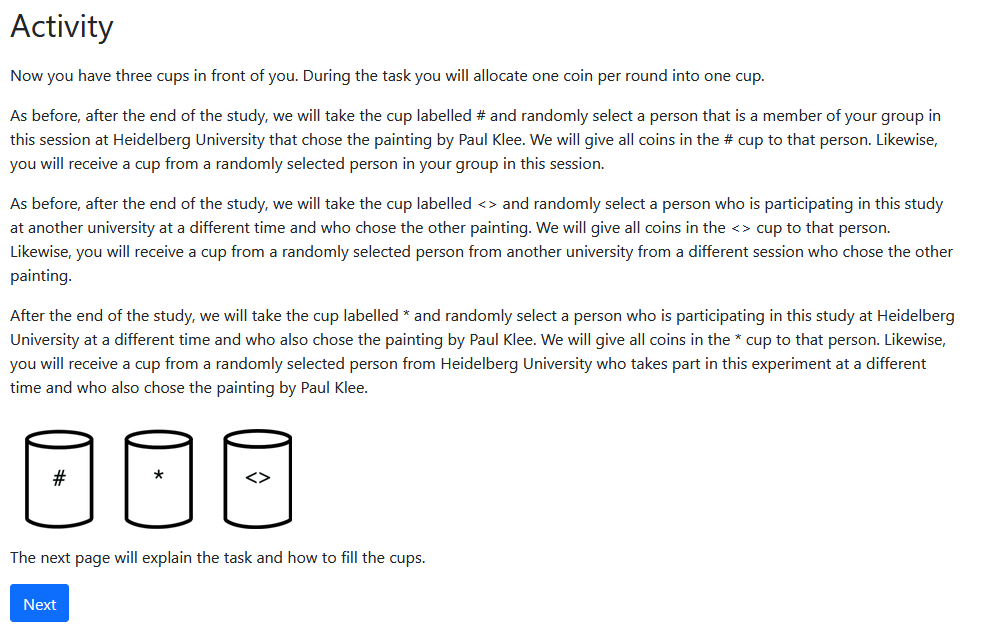}
    \caption*{Treatment Instructions (3 Cups) (i)}
\end{minipage}\hfill
\begin{minipage}{0.48\textwidth}
    \includegraphics[width=\linewidth]{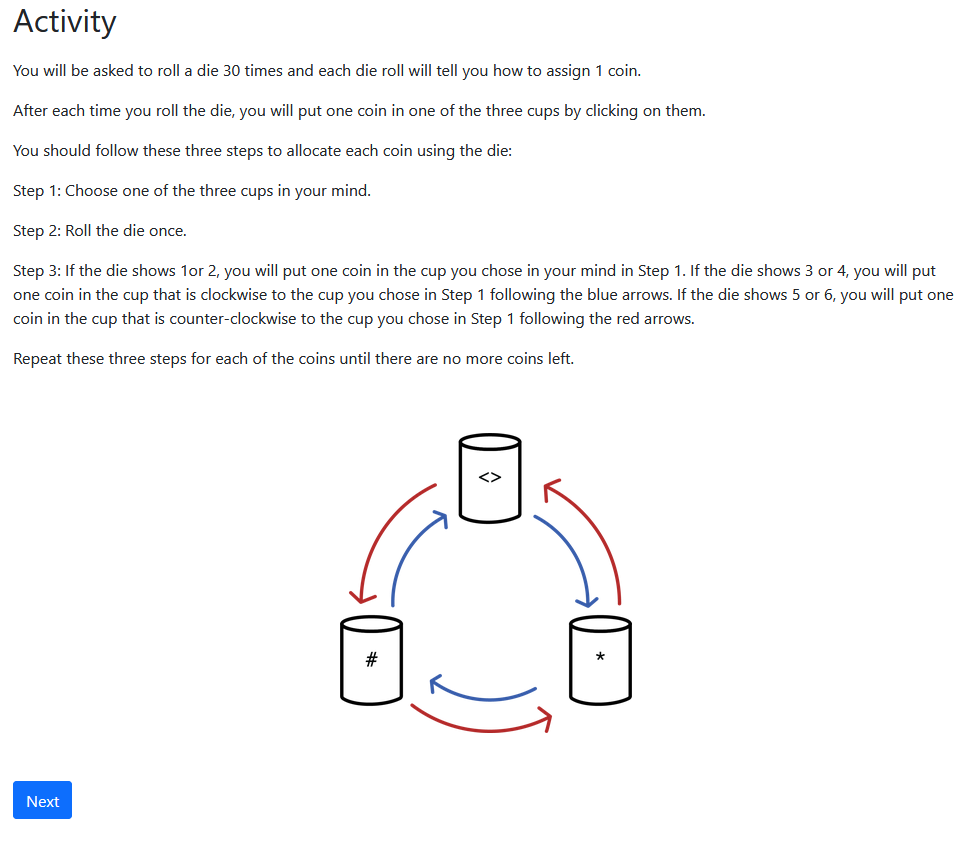}
    \caption*{Treatment Instructions (3 Cups) (ii)}
\end{minipage}

\vspace{1em} 

\begin{minipage}{0.48\textwidth}
    \includegraphics[width=\linewidth]{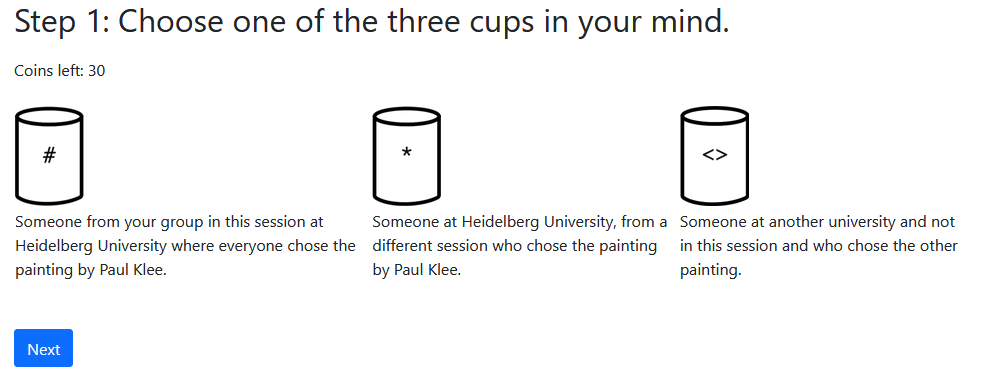}
    \caption*{Round 1 (3 Cups) - Step 1}
\end{minipage}\hfill
\begin{minipage}{0.48\textwidth}
    \includegraphics[width=\linewidth]{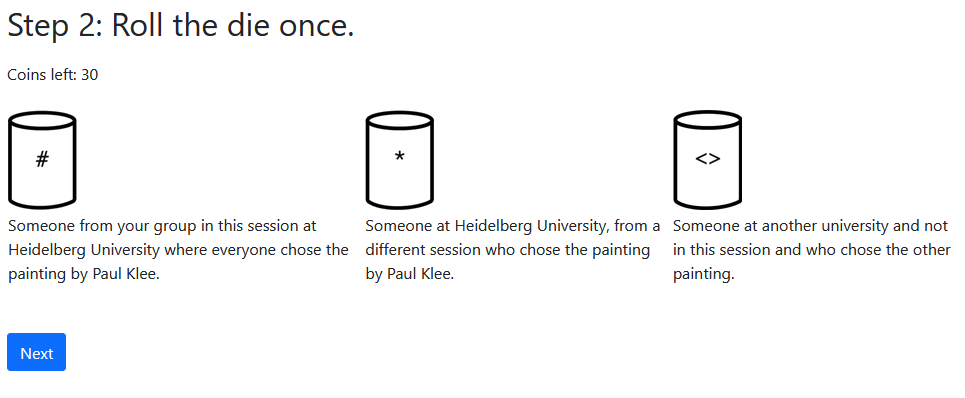}
    \caption*{Round 1 (3 Cups) - Step 2}
\end{minipage}

\vspace{1em} 

\begin{minipage}{0.48\textwidth}
    \centering
    \includegraphics[width=\linewidth]{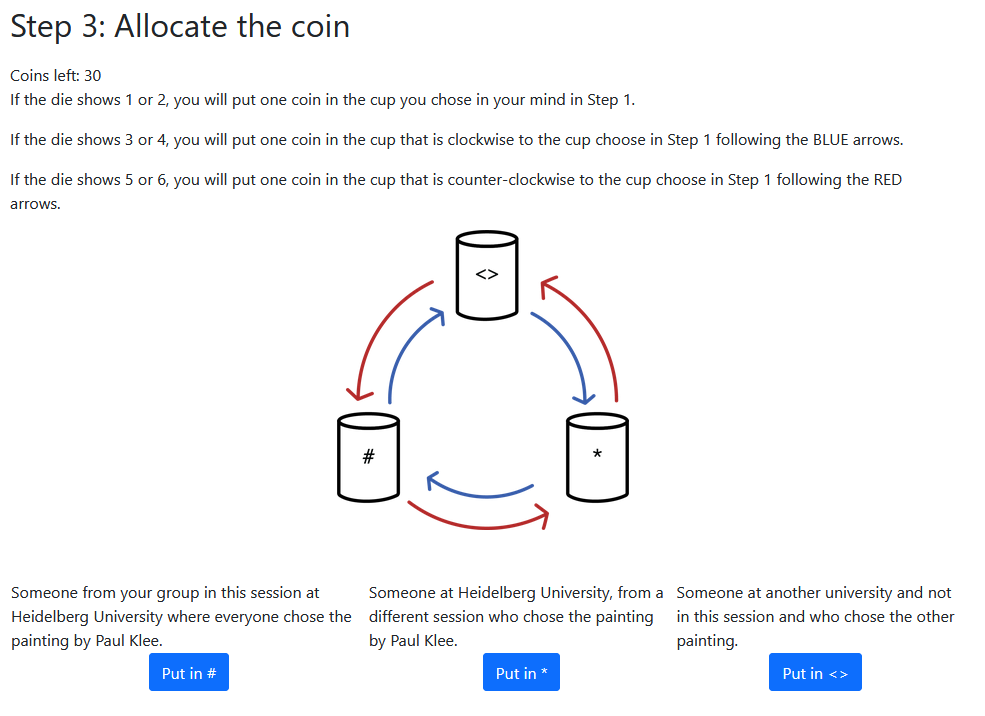}
    \caption*{Round 1 (3 Cups) - Step 3}
\end{minipage}

\caption{Instructions and Task Part 2 (3 cups)}\label{instr-part2}
\end{figure}

 The instructions are shown in figure \ref{instr-part2}.

 \subsection*{Survey and End}

 \begin{enumerate}
     \item \textbf{End of task}
     \item \textbf{Demographic Questionnaire:} Age, Gender, Nationality, Field of Study, Experience with Experiments
     \item \textbf{Payment:} Creation of anonymous payment token (second letter of mothers birthname etc.), forward to separated survey to give personal details for bank transfer.
     \textbf{End}
 \end{enumerate}

\section{All Coin Distributions}

\begin{figure}[ht!]
    \centering
    \includegraphics[width=1\linewidth]{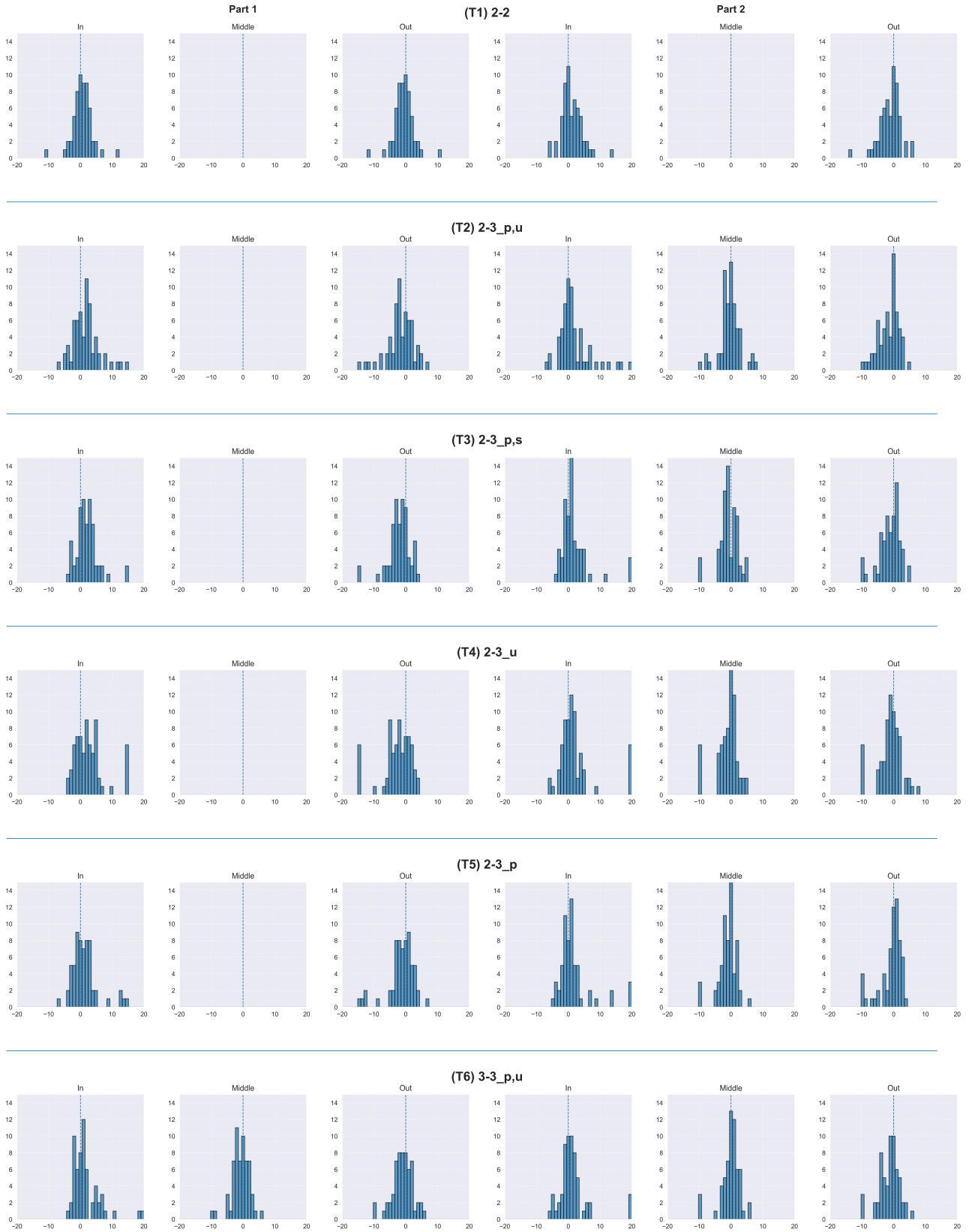}
    \caption{Absolute coin distribution for all cups and treatments. Each row corresponds to one treatment. Within each row, six panels are shown: the left block (Part 1) plots abs\_delta\_first\_in, abs\_delta\_first\_middle, and abs\_delta\_first\_out; the right block (Part 2) plots abs\_delta\_second\_in, abs\_delta\_second\_middle, and abs\_delta\_second\_out. The dashed line at 0 shows the theoretical benchmark.}
    \label{all-coins}
\end{figure}

Figure \ref{all-coins} shows the absolute distribution of all coins in all treatments.

\section{Statistical Results }\label{app:stats}

\subsection{$\%\Delta coins$ against zero benchmark}

In this appendix, we report the full results of the one-sample, two-sided Wilcoxon signed-rank tests used in our main analysis in table \ref{allcoin-data}. For each treatment and outcome variable ($\%\Delta coins$ allocated to the in-group, middle-group, and out-group in Parts 1 and 2), we test whether the median allocation differs from the theoretical expected-value benchmark of zero. This nonparametric test is appropriate for our paired design, where each participant’s allocation is compared against their benchmark value of zero, and it does not require the data to be normally distributed. The p-values reported here correspond to the significance stars shown in figure \ref{dcoin-all}.

\begin{table}[h!]
\centering
\begin{tabular}{c|cccc}

\textbf{Treatment}                    & \multicolumn{1}{c}{\textbf{Part}} & \multicolumn{1}{c}{\textbf{Group}} & \multicolumn{1}{c}{\textbf{Mean (\%)}} & \multicolumn{1}{c}{\textbf{P-Value (Wilcoxon)}}    \\ \cline{1-5}
\multirow{6}{*}{\textbf{(T1) 2-2}}    & \multirow{3}{*}{1}                 & In                                  & 3.8                              & 0.09192                                            \\
                                      &                                    & Middle                              &                                    &                                                    \\
                                      &                                    & Out                                 & -3.8                             & 0.08089                                            \\ \cline{2-5}
                                      & \multirow{3}{*}{2}                 & In                                  & 7.8                              & 0.00937$^{**}$                                     \\
                                      &                                    & Middle                              &                                    &                                                    \\
                                      &                                    & Out                                 & -7.8                             & 0.00609$^{**}$                                     \\ \cline{1-5}
\multirow{6}{*}{\textbf{(T2) 2-3$_{P,U}$}}    & \multirow{3}{*}{1}                 & In                          & 11.2                              & 0.00353$^{**}$                                     \\
                                      &                                    & Middle                              &                                    &                                                    \\
                                      &                                    & Out                                 & -11.2                             & 0.00243$^{**}$                                     \\ \cline{2-5}
                                      & \multirow{3}{*}{2}                 & In                                  & 17.8                              & 0.00903$^{**}$                                     \\
                                      &                                    & Middle                              & -2.9                             & 0.65152                                            \\
                                      &                                    & Out                                 & -14.9                             & 0.00150$^{**}$                                     \\ \cline{1-5}
\multirow{6}{*}{\textbf{(T3) 2-3$_{P,S}$}}    & \multirow{3}{*}{1}         & In                                  &  13.9                              &  0.00002$^{***}$                                                  \\
                                      &                                    & Middle                              &                                    &                                                    \\
                                      &                                    & Out                                 &  -13.9                             &  0.00002$^{***}$                                                   \\ \cline{2-5}
                                      & \multirow{3}{*}{2}                 & In                                  &  19.0                              & 0.00029$^{***}$                                                   \\
                                      &                                    & Middle                              & -7.5                                   & 0.20799                                                   \\
                                      &                                    & Out                                 &  -11.6                                  &  0.03199$^{*}$                                                  \\ \cline{1-5}
\multirow{6}{*}{\textbf{(T4) 2-3$_{U}$}}& \multirow{3}{*}{1}               & In                                  &  18.3                              & 0.00001$^{***}$                                                    \\
                                      &                                    & Middle                              &                                    &                                                    \\
                                      &                                    & Out                                 & -18.3                                   & 0.00001$^{***}$                                                   \\ \cline{2-5}
                                      & \multirow{3}{*}{2}                 & In                                  &  22.5                                  & 0.00198$^{**}$                                                    \\
                                      &                                    & Middle                              & -11.0                                   & 0.06023                                                   \\
                                      &                                    & Out                                 & -11.4                                   & 0.04264$^{*}$                                                   \\ \cline{1-5}
\multirow{6}{*}{\textbf{(T5) 2-3$_{P}$}} & \multirow{3}{*}{1}                 & In                                  & 8.6                              & 0.04527$^{*}$                                     \\
                                      &                                    & Middle                              &                                    &                                                    \\
                                      &                                    & Out                                 & -8.6                             & 0.04167$^{*}$                                      \\ \cline{2-5}
                                      & \multirow{3}{*}{2}                 & In                                  & 17.6                              & 0.01746$^{*}$                                      \\
                                      &                                    & Middle                              & -10.5                             & 0.01322$^{*}$                                      \\
                                      &                                    & Out                                 & -7.1                             & 0.98066                                            \\ \cline{1-5}

\multirow{6}{*}{\textbf{(T6) 3-3$_{P,U}$}}    & \multirow{3}{*}{1}                 & In                          & 17.8                              & 0.00731$^{**}$                                     \\
                                      &                                    & Middle                              & -8.0                             & 0.06122                                            \\
                                      &                                    & Out                                 & -9.8                             & 0.04672$^{*}$                                      \\ \cline{2-5}
                                      & \multirow{3}{*}{2}                 & In                                  & 14.3                              & 0.01674$^{*}$                                      \\
                                      &                                    & Middle                              & -1.0                             & 0.34380                                            \\
                                      &                                    & Out                                 & -13.5                             & 0.00682$^{**}$                                     \\ \cline{1-5}
\end{tabular}
\caption{One-sample, two-sided Wilcoxon signed-rank test results for $\% \Delta coins$ by treatment, part, and group. Reported are group means, and corresponding p-values indicate significance relative to the expected-value benchmark of zero ($*$ for $p<0.05$, $**$ for $p<0.01$, $***$ for $p<0.001$). Data match those shown in Figure \ref{dcoin-all}.}\label{allcoin-data}
\end{table}

Note for table \ref{allcoin-data}: Within each treatment–part combination, mean allocations across the three groups sum to zero by the design of the task (zero-sum allocation). As a result, the Wilcoxon signed-rank tests for each group are not statistically independent, and their p-values are related but not exact inverses of one another. This dependence does not invalidate their interpretation here, as each test addresses a distinct and substantively meaningful question—whether the median allocation to that group differs from the expected-value benchmark of zero. Because the zero-sum constraint couples the outcomes, these tests are not part of a large, independent family of hypotheses, and a multiple-comparison adjustment is not required for our purposes. The p-values should thus be read as direct evidence for or against deviation from a rule-following allocation behaviour towards each group.

\subsection{$\%\Delta coins$ from part 1 to 2}

We compare allocations between Part 1 and Part 2 within each group (In, Middle, Out) using the Wilcoxon signed-rank test for paired data. This nonparametric test evaluates whether the median change from Part 1 to Part 2 differs from zero for each treatment separately. The output includes the number of usable pairs (i.e., non-zero changes), the median change, the Wilcoxon W statistic, and the two-sided p-value.

\begin{table}[h!]
\centering
\begin{tabular}{c|ccccc}
\textbf{Treatment}           & \textbf{Group}  & \textbf{N (pairs)}  & \textbf{Median change (pct points)} & \textbf{Wilcoxon W} & \textbf{P-Value}  \\ \hline
\multirow{2}{*}{\textbf{(T1) 2-2}}    & In     & 49        & 6.67      & 497.0      & 0.255137 \\
                             & Out    & 49        & -6.67     & 494.0      & 0.242851 \\ \hline
\multirow{2}{*}{\textbf{(T2) 2-3$_{P,U}$}}    & In     & 53        & 3.33      & 624.5      & 0.420329 \\
                             & Out    & 55        & 3.33      & 702.5      & 0.571551 \\ \hline
\multirow{2}{*}{\textbf{(T3) 2-3$_{P,S}$}}    & In     &  60       & $\approx 0$          & 873.5      & 0.759873 \\
                             & Out    & 54        & 3.33      & 655.5      & 0.453567  \\ \hline
\multirow{2}{*}{\textbf{(T4) 2-3$_{U}$}}    & In     &  69       &  3.33         & 1170.0      & 0.822484 \\
                             & Out    &  61       & 3.33      & 713.5      & 0.095474  \\ \hline
\multirow{2}{*}{\textbf{(T5) 2-3$_P$}} & In     & 61        & 6.67      & 719.5      & 0.104386 \\
                             & Out    & 59        & 3.33      & 787.0      & 0.459121 \\ \hline
\multirow{3}{*}{\textbf{(T6) 3-3$_{P,U}$}}    & In     & 47        & -10.00     & 436.0      & 0.178929 \\
                             & Middle & 52        & 10.00      & 452.5      & 0.030816$^{*}$ \\
                             & Out    & 54        & -10.00     & 699.0      & 0.707502
\end{tabular}
\caption{Wilcoxon signed-rank test results comparing allocations between Part 1 and Part 2 within each treatment and group (In, Middle, Out). For each group, the test evaluates whether the median change in $\%\Delta coins$ from Part 1 to Part 2 differs from zero. “N (pairs)” indicates the number of usable paired observations (excluding cases with zero change). Median change is reported in $\%\Delta coins$; Wilcoxon W is the test statistic; p-values are two-sided.}
\end{table}

\subsection{OLS Regression \textbf{$\%\Delta coins$ middle-group part 2} With Controls}

{\small
\begin{longtable}{@{}lcccccc@{}}
\caption{OLS Regression Results on \textbf{$\%\Delta coins$ middle-group part 2} with demographic controls. Each treatment coefficient represents the conditional mean $\%\Delta coins$ for that treatment's middle-group (at reference levels of control variables). Model estimated without intercept.}\label{tab:delta-second-middle-controls}\\
\toprule
                            & Coef.   & Std. Err. & t       & $P>|t|$   & [0.025   & 0.975] \\
\midrule
\endfirsthead

\multicolumn{7}{l}{\small\textit{Table \ref{tab:delta-second-middle-controls} continued}}\\
\toprule
                            & Coef.   & Std. Err. & t       & $P>|t|$   & [0.025   & 0.975] \\
\midrule
\endhead

\midrule
\multicolumn{7}{r}{\textit{Continued on next page}}\\
\endfoot

\bottomrule
\endlastfoot

\multicolumn{7}{@{}l}{\textbf{Treatment Variables}}\\
(T2) 2-3$_{P,U}$            & 0.0115  & 0.301     & 0.038   & 0.970   & -0.583   & 0.606 \\
(T3) 2-3$_{P,S}$            & -0.0144 & 0.305     & -0.047  & 0.962   & -0.617   & 0.588 \\
(T4) 2-3$_{U}$              & 0.0030  & 0.304     & 0.010   & 0.992   & -0.597   & 0.603 \\
(T5) 2-3$_{P}$              & -0.0503 & 0.307     & -0.164  & 0.870   & -0.656   & 0.555 \\
\midrule
\multicolumn{7}{@{}l}{\textbf{Gender (ref: Female)}}\\
Male                        & 0.0137  & 0.051     & 0.267   & 0.790   & -0.088   & 0.115 \\
Non-binary                  & -0.0661 & 0.330     & -0.200  & 0.841   & -0.718   & 0.585 \\
Prefer not to tell          & -0.1032 & 0.355     & -0.291  & 0.771   & -0.804   & 0.597 \\
\midrule
\multicolumn{7}{@{}l}{\textbf{Nationality (ref: first alphabetically)}}\\
Bangladesh                  & 0.4089  & 0.384     & 1.065   & 0.288   & -0.349   & 1.167 \\
Bosnia and Herzegovina      & 0.1986  & 0.386     & 0.515   & 0.607   & -0.563   & 0.960 \\
Brazil                      & 0.1320  & 0.373     & 0.353   & 0.724   & -0.605   & 0.869 \\
Bulgaria                    & 0.3056  & 0.301     & 1.015   & 0.312   & -0.289   & 0.900 \\
China                       & 0.1796  & 0.239     & 0.752   & 0.453   & -0.292   & 0.651 \\
Colombia                    & -0.1450 & 0.354     & -0.410  & 0.682   & -0.843   & 0.553 \\
Cyprus                      & 0.6737  & 0.410     & 1.642   & 0.103   & -0.136   & 1.484 \\
Ecuador                     & -0.6170 & 0.290     & -2.126  & 0.035   & -1.190   & -0.044 \\
Egypt                       & 0.5960  & 0.279     & 2.137   & 0.034   & 0.045    & 1.147 \\
Georgia                     & 0.4286  & 0.383     & 1.120   & 0.264   & -0.327   & 1.184 \\
Germany                     & 0.1944  & 0.208     & 0.933   & 0.352   & -0.217   & 0.606 \\
Guatemala                   & -0.0012 & 0.395     & -0.003  & 0.998   & -0.780   & 0.778 \\
India                       & 0.2740  & 0.215     & 1.273   & 0.205   & -0.151   & 0.699 \\
Indonesia                   & 0.2095  & 0.306     & 0.686   & 0.494   & -0.394   & 0.813 \\
Iran                        & 0.2827  & 0.257     & 1.098   & 0.274   & -0.225   & 0.791 \\
Italy                       & 0.1270  & 0.306     & 0.415   & 0.679   & -0.477   & 0.731 \\
Japan                       & 0.8321  & 0.308     & 2.699   & 0.008   & 0.224    & 1.441 \\
Kazakhstan                  & 0.2052  & 0.302     & 0.679   & 0.498   & -0.391   & 0.802 \\
Kyrgyzstan                  & -0.0361 & 0.414     & -0.087  & 0.931   & -0.853   & 0.780 \\
Latvia                      & 0.4016  & 0.278     & 1.443   & 0.151   & -0.148   & 0.951 \\
Moldova                     & 0.0535  & 0.373     & 0.144   & 0.886   & -0.682   & 0.789 \\
Myanmar                     & 0.1631  & 0.389     & 0.419   & 0.676   & -0.605   & 0.931 \\
Nepal                       & -0.3113 & 0.311     & -1.001  & 0.318   & -0.925   & 0.303 \\
Nigeria                     & 0.1235  & 0.293     & 0.422   & 0.674   & -0.454   & 0.701 \\
Pakistan                    & 0.4621  & 0.252     & 1.837   & 0.068   & -0.035   & 0.959 \\
Peru                        & 0.6868  & 0.372     & 1.845   & 0.067   & -0.048   & 1.422 \\
Philippines                 & -0.0322 & 0.393     & -0.082  & 0.935   & -0.807   & 0.743 \\
Poland                      & 0.5403  & 0.372     & 1.453   & 0.148   & -0.194   & 1.274 \\
Portugal                    & 0.3233  & 0.353     & 0.916   & 0.361   & -0.373   & 1.020 \\
Romania                     & 0.0761  & 0.281     & 0.271   & 0.786   & -0.478   & 0.630 \\
Russia                      & 0.2374  & 0.276     & 0.859   & 0.392   & -0.308   & 0.783 \\
Sierra Leone                & 0.4200  & 0.418     & 1.006   & 0.316   & -0.404   & 1.244 \\
South Korea                 & 0.1001  & 0.304     & 0.329   & 0.742   & -0.500   & 0.701 \\
Spain                       & 0.0876  & 0.383     & 0.229   & 0.820   & -0.669   & 0.844 \\
Togo                        & 0.0162  & 0.357     & 0.045   & 0.964   & -0.689   & 0.722 \\
Turkey                      & -0.0106 & 0.259     & -0.041  & 0.967   & -0.523   & 0.502 \\
Ukraine                     & 0.0339  & 0.312     & 0.109   & 0.914   & -0.583   & 0.651 \\
United Kingdom              & 0.4613  & 0.384     & 1.200   & 0.232   & -0.297   & 1.220 \\
United States               & 0.8759  & 0.382     & 2.291   & 0.023   & 0.121    & 1.631 \\
Vietnam                     & 0.3669  & 0.234     & 1.570   & 0.118   & -0.094   & 0.828 \\
Zimbabwe                    & 0.3053  & 0.365     & 0.837   & 0.404   & -0.414   & 1.025 \\
\midrule
\multicolumn{7}{@{}l}{\textbf{Field of Study (ref: first alphabetically)}}\\
Architecture \& Design      & 0.2284  & 0.402     & 0.568   & 0.571   & -0.565   & 1.022 \\
Arts                        & 0.1921  & 0.358     & 0.536   & 0.592   & -0.515   & 0.899 \\
Biology                     & -0.0501 & 0.209     & -0.240  & 0.811   & -0.463   & 0.362 \\
Business                    & 0.0328  & 0.207     & 0.158   & 0.874   & -0.376   & 0.442 \\
Chemistry                   & 0.0420  & 0.214     & 0.196   & 0.845   & -0.381   & 0.465 \\
Computer Science            & -0.1826 & 0.211     & -0.864  & 0.389   & -0.600   & 0.234 \\
Earth Science               & 0.3664  & 0.419     & 0.875   & 0.383   & -0.460   & 1.193 \\
Economics                   & 0.0115  & 0.183     & 0.063   & 0.950   & -0.350   & 0.373 \\
Education / Teaching        & -0.2302 & 0.221     & -1.043  & 0.298   & -0.666   & 0.205 \\
Engineering \& Technology   & -0.0223 & 0.194     & -0.115  & 0.909   & -0.405   & 0.361 \\
Environmental Studies       & 0.1617  & 0.353     & 0.458   & 0.647   & -0.535   & 0.858 \\
Geography                   & 0.0411  & 0.355     & 0.116   & 0.908   & -0.661   & 0.743 \\
History                     & 0.2168  & 0.357     & 0.606   & 0.545   & -0.489   & 0.922 \\
Journalism \& Media         & -0.8927 & 0.355     & -2.512  & 0.013   & -1.594   & -0.191 \\
Languages \& Literature     & 0.0727  & 0.203     & 0.358   & 0.721   & -0.328   & 0.473 \\
Law                         & -0.0394 & 0.200     & -0.197  & 0.844   & -0.434   & 0.355 \\
Mathematics                 & 0.1687  & 0.252     & 0.669   & 0.505   & -0.329   & 0.666 \\
Medicine                    & -0.0562 & 0.207     & -0.272  & 0.786   & -0.465   & 0.352 \\
Not a student               & 0.0988  & 0.220     & 0.449   & 0.654   & -0.335   & 0.532 \\
Other                       & 0.1587  & 0.207     & 0.767   & 0.444   & -0.250   & 0.567 \\
Philosophy                  & 0.1250  & 0.253     & 0.495   & 0.621   & -0.374   & 0.624 \\
Physics                     & -0.0062 & 0.204     & -0.030  & 0.976   & -0.409   & 0.397 \\
Political Science           & 0.0576  & 0.197     & 0.293   & 0.770   & -0.331   & 0.446 \\
Psychology                  & 0.7061  & 0.351     & 2.013   & 0.046   & 0.014    & 1.399 \\
Social Work                 & 0.0532  & 0.351     & 0.152   & 0.880   & -0.639   & 0.745 \\
Sociology                   & 0.0338  & 0.214     & 0.158   & 0.875   & -0.388   & 0.456 \\
\midrule
\multicolumn{7}{@{}l}{\textbf{Other Controls}}\\
Experience (ordinal)        & -0.0542 & 0.022     & -2.502  & 0.013   & -0.097   & -0.011 \\
Art preference              & -0.0312 & 0.048     & -0.654  & 0.514   & -0.125   & 0.063 \\
Age                         & -0.0079 & 0.005     & -1.570  & 0.118   & -0.018   & 0.002 \\
\midrule
\multicolumn{7}{@{}l}{\textbf{Model Information:}}\\
\multicolumn{2}{l}{R-squared}          & \multicolumn{1}{r}{0.396} & \multicolumn{4}{r}{}\\
\multicolumn{2}{l}{Adj. R-squared}     & \multicolumn{1}{r}{0.126} & \multicolumn{2}{l}{No. Observations}& \multicolumn{1}{r}{247} & \\
\multicolumn{2}{l}{F-statistic}        & \multicolumn{1}{r}{1.468} & \multicolumn{2}{l}{Df Residuals}    & \multicolumn{1}{r}{170} & \\
\multicolumn{2}{l}{Prob (F-statistic)} & \multicolumn{1}{r}{0.021} & \multicolumn{2}{l}{Df Model}        & \multicolumn{1}{r}{76} & \\
\multicolumn{2}{l}{Log-Likelihood}     & \multicolumn{1}{r}{-6.395} & \multicolumn{2}{l}{Covariance Type} & \multicolumn{1}{r}{Nonrobust} & \\
\end{longtable}
}

\section{Additional Analyses}\label{pre-reg}
The main analysis above contains all important aspects to draw insights from our data set. The following analyses are exactly as pre registered (\url{https://www.socialscienceregistry.org/trials/9670}). In hindsight, some of these approaches are unnecessarily complicated and the analysis above are much simpler and to the point. Other approaches do not yield statistically significant results due to the highly stochastic nature of the task which makes it hard to detect small effect sizes given the sample size. In hindsight, this could have also been partially expected. 

\subsection*{(4) Causal effect of introducing a middle-group:}

\textit{Original text: We will use a difference-in-difference approach to identify the causal effect of introducing a middle-group. The control group is the “2-2” group, the treatment group the “2-3” group. The first part is the pre-treatment, the second part is the post-treatment period. We will look at two outcome measures correcting for multiple hypothesis testing: (i) When a middle-group is introduced, what is the effect on the delta in allocation of coins compared to the theoretical value for the in-group? (ii) … for the out-group? $H_0$: there are no effects.}

In the regression results in table \ref{4-in} and \ref{4-out} we cannot detect any statistically significant effect for neither in- nor out-group. 

\begin{table}[ht!]
\centering

\begin{tabular}{@{}lcccccc@{}}
\toprule
                            & Coef.   & Std. Err. & t       & $P>|t|$   & [0.025   & 0.975] \\
\midrule
Intercept                   & 0.0384  & 0.043     & 0.886   & 0.377   & -0.047   & 0.124  \\
Treatment Group                       & 0.0738  & 0.060     & 1.222   & 0.223   & -0.045   & 0.193  \\
Post Treatment                        & 0.0395  & 0.061     & 0.645   & 0.520   & -0.081   & 0.160  \\
Interaction [Treatment x Post]                & 0.0261  & 0.085     & 0.305   & 0.760   & -0.142   & 0.194  \\
\midrule
\multicolumn{7}{@{}l}{\textbf{Model Information:}}\\
\multicolumn{2}{l}{R-squared}       & \multicolumn{1}{r}{0.023} & \multicolumn{4}{r}{}\\
\multicolumn{2}{l}{Adj. R-squared}  & \multicolumn{1}{r}{0.011} & \multicolumn{2}{l}{No. Observations}& \multicolumn{1}{r}{244} & \multicolumn{1}{r}{}\\
\multicolumn{2}{l}{F-statistic}     & \multicolumn{1}{r}{1.925} & \multicolumn{2}{l}{Df Residuals}    & \multicolumn{1}{r}{240} & \multicolumn{1}{r}{}\\
\multicolumn{2}{l}{Prob (F-statistic)} & \multicolumn{1}{r}{0.126} & \multicolumn{2}{l}{Df Model}        & \multicolumn{1}{r}{3} & \multicolumn{1}{r}{}\\
\multicolumn{2}{l}{Log-Likelihood}  & \multicolumn{1}{r}{-75.967} & \multicolumn{2}{l}{Covariance Type} & \multicolumn{1}{r}{Nonrobust} & \multicolumn{1}{r}{}\\
\bottomrule
\end{tabular}
\caption{Difference-in-Differences: Effect of introducing a middle-group on in-group allocations. 
The control group is T1 (2-2), where only in- and out-groups are present throughout. 
The treatment group is T2 (2-3$_{P,U}$), where a middle-group is introduced in Part 2. 
\textit{Intercept}: baseline in-group allocation in T1, Part 1. 
\textit{Treatment Group}: difference between T2 and T1 in Part 1. 
\textit{Post Treatment}: change from Part 1 to Part 2 in T1. 
\textit{Interaction}: the DiD estimator—additional change in T2 relative to T1 when the middle-group is introduced. }\label{4-in}
\end{table}

\begin{table}[ht!]
\centering

\begin{tabular}{@{}lcccccc@{}}
\toprule
                            & Coef.   & Std. Err. & t       & $P>|t|$   & [0.025   & 0.975] \\
\midrule
Intercept                   & -0.0384 & 0.034     & -1.118  & 0.265   & -0.106   & 0.029  \\
Treatment Group             & -0.0738 & 0.048     & -1.542  & 0.124   & -0.168   & 0.020  \\
Post Treatment              & -0.0395 & 0.049     & -0.814  & 0.417   & -0.135   & 0.056  \\
Interaction [Treatment x Post] & 0.0025  & 0.068     & 0.037   & 0.970   & -0.131   & 0.136  \\
\midrule
\multicolumn{7}{@{}l}{\textbf{Model Information:}}\\
\multicolumn{2}{l}{R-squared}       & \multicolumn{1}{r}{0.024} & \multicolumn{4}{r}{}\\
\multicolumn{2}{l}{Adj. R-squared}  & \multicolumn{1}{r}{0.012} & \multicolumn{2}{l}{No. Observations}& \multicolumn{1}{r}{244} & \multicolumn{1}{r}{}\\
\multicolumn{2}{l}{F-statistic}     & \multicolumn{1}{r}{1.960} & \multicolumn{2}{l}{Df Residuals}    & \multicolumn{1}{r}{240} & \multicolumn{1}{r}{}\\
\multicolumn{2}{l}{Prob (F-statistic)} & \multicolumn{1}{r}{0.121} & \multicolumn{2}{l}{Df Model}        & \multicolumn{1}{r}{3} & \multicolumn{1}{r}{}\\
\multicolumn{2}{l}{Log-Likelihood}  & \multicolumn{1}{r}{-19.169} & \multicolumn{2}{l}{Covariance Type} & \multicolumn{1}{r}{Nonrobust} & \multicolumn{1}{r}{}\\
\bottomrule
\end{tabular}
\caption{Difference-in-Differences: Effect of introducing a middle-group on out-group allocations. 
The control group is T1 (2-2), the treatment group is T2 (2-3$_{P,U}$). 
\textit{Interaction}: the DiD estimator captures whether introducing a middle-group changes allocations to the out-group beyond any time trend.}\label{4-out}
\end{table}

\subsection*{(5) Compare this effect between the alternative middle-groups:}

\textit{Original text: Treatments “2-3” and “2-3a” use two different types of middle-groups. Everything we do for the “2-3” treatment in the last part (4) of the analysis we will also do for treatment “2-3a” and then compare the coefficients and their standard deviations.}

In the regression results in table \ref{5-in} and \ref{5-out} we cannot detect any statistically significant effect for neither in- nor out-group. 

\begin{table}[ht!]
\centering

\begin{tabular}{@{}lcccccc@{}}
\toprule
                            & Coef.   & Std. Err. & t       & $P>|t|$   & [0.025   & 0.975] \\
\midrule
Intercept                   & 0.0384  & 0.045     & 0.857   & 0.392   & -0.050   & 0.127  \\
Treatment Group             & 0.0476  & 0.063     & 0.760   & 0.448   & -0.076   & 0.171  \\
Post Treatment              & 0.0395  & 0.063     & 0.624   & 0.533   & -0.085   & 0.164  \\
Interaction [Treatment x Post] & 0.0502  & 0.089     & 0.567   & 0.571   & -0.124   & 0.225  \\
\midrule
\multicolumn{7}{@{}l}{\textbf{Model Information:}}\\
\multicolumn{2}{l}{R-squared}       & \multicolumn{1}{r}{0.021} & \multicolumn{4}{r}{}\\
\multicolumn{2}{l}{Adj. R-squared}  & \multicolumn{1}{r}{0.009} & \multicolumn{2}{l}{No. Observations}& \multicolumn{1}{r}{242} & \multicolumn{1}{r}{}\\
\multicolumn{2}{l}{F-statistic}     & \multicolumn{1}{r}{1.730} & \multicolumn{2}{l}{Df Residuals}    & \multicolumn{1}{r}{238} & \multicolumn{1}{r}{}\\
\multicolumn{2}{l}{Prob (F-statistic)} & \multicolumn{1}{r}{0.161} & \multicolumn{2}{l}{Df Model}        & \multicolumn{1}{r}{3} & \multicolumn{1}{r}{}\\
\multicolumn{2}{l}{Log-Likelihood}  & \multicolumn{1}{r}{-83.433} & \multicolumn{2}{l}{Covariance Type} & \multicolumn{1}{r}{Nonrobust} & \multicolumn{1}{r}{}\\
\bottomrule
\end{tabular}
\caption{Difference-in-Differences: Comparing alternative middle-groups—effect on in-group allocations. 
The control group is T2 (2-3$_{P,U}$), where the middle-group shares painting preference and university affiliation. 
The treatment group is T5 (2-3$_{P}$), where the middle-group shares only painting preference. 
\textit{Interaction}: the DiD estimator captures whether the type of middle-group differentially affects in-group allocations.}\label{5-in}
\end{table}

\begin{table}[ht!]
\centering

\begin{tabular}{@{}lcccccc@{}}
\toprule
                            & Coef.   & Std. Err. & t       & $P>|t|$   & [0.025   & 0.975] \\
\midrule
Intercept                   & -0.0384 & 0.036     & -1.079  & 0.282   & -0.109   & 0.032  \\
Treatment Group             & -0.0476 & 0.050     & -0.957  & 0.339   & -0.146   & 0.050  \\
Post Treatment              & -0.0395 & 0.050     & -0.786  & 0.433   & -0.139   & 0.060  \\
Interaction [Treatment x Post] & 0.0546  & 0.070     & 0.776   & 0.438   & -0.084   & 0.193  \\
\midrule
\multicolumn{7}{@{}l}{\textbf{Model Information:}}\\
\multicolumn{2}{l}{R-squared}       & \multicolumn{1}{r}{0.004} & \multicolumn{4}{r}{}\\
\multicolumn{2}{l}{Adj. R-squared}  & \multicolumn{1}{r}{-0.008} & \multicolumn{2}{l}{No. Observations}& \multicolumn{1}{r}{242} & \multicolumn{1}{r}{}\\
\multicolumn{2}{l}{F-statistic}     & \multicolumn{1}{r}{0.3482} & \multicolumn{2}{l}{Df Residuals}    & \multicolumn{1}{r}{238} & \multicolumn{1}{r}{}\\
\multicolumn{2}{l}{Prob (F-statistic)} & \multicolumn{1}{r}{0.790} & \multicolumn{2}{l}{Df Model}        & \multicolumn{1}{r}{3} & \multicolumn{1}{r}{}\\
\multicolumn{2}{l}{Log-Likelihood}  & \multicolumn{1}{r}{-27.536} & \multicolumn{2}{l}{Covariance Type} & \multicolumn{1}{r}{Nonrobust} & \multicolumn{1}{r}{}\\
\bottomrule
\end{tabular}
\caption{Difference-in-Differences: Comparing alternative middle-groups—effect on out-group allocations. 
The control group is T2 (2-3$_{P,U}$), the treatment group is T5 (2-3$_{P}$). 
\textit{Interaction}: the DiD estimator captures whether the type of middle-group differentially affects out-group allocations.}\label{5-out}
\end{table}

\subsection*{(6) Effect of having had a middle-group from the beginning versus being introduced to it later:}

\textit{Original text: The treatment “3-3” had the middle-group from the start, the treatment “2-3” only in the second part. To investigate the effect of this difference in experience with a middle-group, we will pool all data from the second part of these two treatments and run an OLS regression with the treatment-variable (i.e. the difference in experience) as explanatory variable. We will do this for three outcome measures correcting for multiple hypothesis testing: (i) What is the effect of being introduced to a middle-group already at the start on the delta in allocation of coins compared to the theoretical value for the in-group? (ii) … for the middle-group? (iii) … for the out-group? $H_0$: There are no effects.}

In the regression results in table \ref{6-in}, \ref{6-middle} and \ref{6-out} we cannot detect any statistically significant effect for neither in- nor, middle-, nor out-group. 

\begin{table}[ht!]
\centering

\begin{tabular}{@{}lcccccc@{}}
\toprule
                            & Coef.   & Std. Err. & t       & $P>|t|$   & [0.025   & 0.975] \\
\midrule
Intercept                   & 0.1778  & 0.064     & 2.764   & 0.007   & 0.050    & 0.305  \\
Treatment Dummy             & -0.0344 & 0.092     & -0.374  & 0.709   & -0.217   & 0.148  \\
\midrule
\multicolumn{7}{@{}l}{\textbf{Model Information:}}\\
\multicolumn{2}{l}{R-squared}       & \multicolumn{1}{r}{0.001} & \multicolumn{4}{r}{}\\
\multicolumn{2}{l}{Adj. R-squared}  & \multicolumn{1}{r}{-0.007} & \multicolumn{2}{l}{No. Observations}& \multicolumn{1}{r}{123} & \multicolumn{1}{r}{}\\
\multicolumn{2}{l}{F-statistic}     & \multicolumn{1}{r}{0.1399} & \multicolumn{2}{l}{Df Residuals}    & \multicolumn{1}{r}{121} & \multicolumn{1}{r}{}\\
\multicolumn{2}{l}{Prob (F-statistic)} & \multicolumn{1}{r}{0.709} & \multicolumn{2}{l}{Df Model}        & \multicolumn{1}{r}{1} & \multicolumn{1}{r}{}\\
\multicolumn{2}{l}{Log-Likelihood}  & \multicolumn{1}{r}{-90.825} & \multicolumn{2}{l}{Covariance Type} & \multicolumn{1}{r}{Nonrobust} & \multicolumn{1}{r}{}\\
\bottomrule
\end{tabular}
\caption{Effect of prior middle-group exposure on in-group allocations (Part 2 only). 
Compares T2 (2-3$_{P,U}$, middle-group introduced in Part 2) with T6 (3-3$_{P,U}$, middle-group present from Part 1). 
\textit{Intercept}: mean \%$\Delta$coins for in-group in T2. 
\textit{Treatment Dummy}: difference for participants who had prior exposure to the middle-group (T6 vs.\ T2).}\label{6-in}
\end{table}

\begin{table}[ht!]
\centering

\begin{tabular}{@{}lcccccc@{}}
\toprule
                            & Coef.   & Std. Err. & t       & $P>|t|$   & [0.025   & 0.975] \\
\midrule
Intercept                   & -0.0286 & 0.040     & -0.716  & 0.475   & -0.108   & 0.050  \\
Treatment Dummy             & 0.0186  & 0.057     & 0.325   & 0.746   & -0.095   & 0.132  \\
\midrule
\multicolumn{7}{@{}l}{\textbf{Model Information:}}\\
\multicolumn{2}{l}{R-squared}       & \multicolumn{1}{r}{0.001} & \multicolumn{4}{r}{}\\
\multicolumn{2}{l}{Adj. R-squared}  & \multicolumn{1}{r}{-0.007} & \multicolumn{2}{l}{No. Observations}& \multicolumn{1}{r}{123} & \multicolumn{1}{r}{}\\
\multicolumn{2}{l}{F-statistic}     & \multicolumn{1}{r}{0.1056} & \multicolumn{2}{l}{Df Residuals}    & \multicolumn{1}{r}{121} & \multicolumn{1}{r}{}\\
\multicolumn{2}{l}{Prob (F-statistic)} & \multicolumn{1}{r}{0.746} & \multicolumn{2}{l}{Df Model}        & \multicolumn{1}{r}{1} & \multicolumn{1}{r}{}\\
\multicolumn{2}{l}{Log-Likelihood}  & \multicolumn{1}{r}{-32.128} & \multicolumn{2}{l}{Covariance Type} & \multicolumn{1}{r}{Nonrobust} & \multicolumn{1}{r}{}\\
\bottomrule
\end{tabular}
\caption{Effect of prior middle-group exposure on middle-group allocations (Part 2 only). 
Compares T2 (2-3$_{P,U}$, middle-group introduced in Part 2) with T6 (3-3$_{P,U}$, middle-group present from Part 1). 
\textit{Intercept}: mean \%$\Delta$coins for middle-group in T2. 
\textit{Treatment Dummy}: difference for participants who had prior exposure to the middle-group (T6 vs.\ T2).}\label{6-middle}
\end{table}

\begin{table}[ht!]
\centering

\begin{tabular}{@{}lcccccc@{}}
\toprule
                            & Coef.   & Std. Err. & t       & $P>|t|$   & [0.025   & 0.975] \\
\midrule
Intercept                   & -0.1492 & 0.041     & -3.668  & 0.000   & -0.230   & -0.069 \\
Treatment Dummy             & 0.0142  & 0.058     & 0.244   & 0.808   & -0.101   & 0.130  \\
\midrule
\multicolumn{7}{@{}l}{\textbf{Model Information:}}\\
\multicolumn{2}{l}{R-squared}       & \multicolumn{1}{r}{0.000} & \multicolumn{4}{r}{}\\
\multicolumn{2}{l}{Adj. R-squared}  & \multicolumn{1}{r}{-0.008} & \multicolumn{2}{l}{No. Observations}& \multicolumn{1}{r}{123} & \multicolumn{1}{r}{}\\
\multicolumn{2}{l}{F-statistic}     & \multicolumn{1}{r}{0.05950} & \multicolumn{2}{l}{Df Residuals}    & \multicolumn{1}{r}{121} & \multicolumn{1}{r}{}\\
\multicolumn{2}{l}{Prob (F-statistic)} & \multicolumn{1}{r}{0.808} & \multicolumn{2}{l}{Df Model}        & \multicolumn{1}{r}{1} & \multicolumn{1}{r}{}\\
\multicolumn{2}{l}{Log-Likelihood}  & \multicolumn{1}{r}{-34.471} & \multicolumn{2}{l}{Covariance Type} & \multicolumn{1}{r}{Nonrobust} & \multicolumn{1}{r}{}\\
\bottomrule
\end{tabular}
\caption{Effect of prior middle-group exposure on out-group allocations (Part 2 only). 
Compares T2 (2-3$_{P,U}$, middle-group introduced in Part 2) with T6 (3-3$_{P,U}$, middle-group present from Part 1). 
\textit{Intercept}: mean \%$\Delta$coins for out-group in T2. 
\textit{Treatment Dummy}: difference for participants who had prior exposure to the middle-group (T6 vs.\ T2).}\label{6-out}
\end{table}

\subsection*{(7) Analysis on the individual level:}

\textit{Original text: For each participant and part we will count how many coins were put into cup x and calculate the “delta in absolute coins” by subtracting the theoretical amount of coins (10 or 15). We will plot the “delta in absolute coins per cup”-distribution of all participants in each treatment, part and cup on a x-scale [-15, +15]. We will then perform a Monte-Carlo-Simulations for each of these plots where the simulated agents indeed distribute the coins randomly as specified by the rules. The difference to the actual experimental data is that there the participants might in fact not have followed the rules. In this case we will observe different distributions. Then for each treatment, part and cup we compare the simulated results with the actual results of the human experimental participants with a Kolmogorov-Smirnov-test. $H_0$: The distributions of the results of the experimental data and the simulated data are statistically not different from each other.}

In hindsight this elaborate simulation process is not needed. Simple statistical testing against a 0 benchmark is sufficient, as has been the core element of the main text. The simulation results therefore do not give us any meaningful insights beyond the ones presented in the paper.

\subsection*{(8) Behaviour over time:}

\textit{Original text: Per treatment and round we calculate the mean contribution per cup. First, we will plot this variable over all rounds. Then we will perform an OLS regression on this variable with rounds as explanatory variable to see whether there are time trends. $H_0$: there are no time trends.}

\begin{figure}[ht!]
    \centering
    \includegraphics[width=1\linewidth]{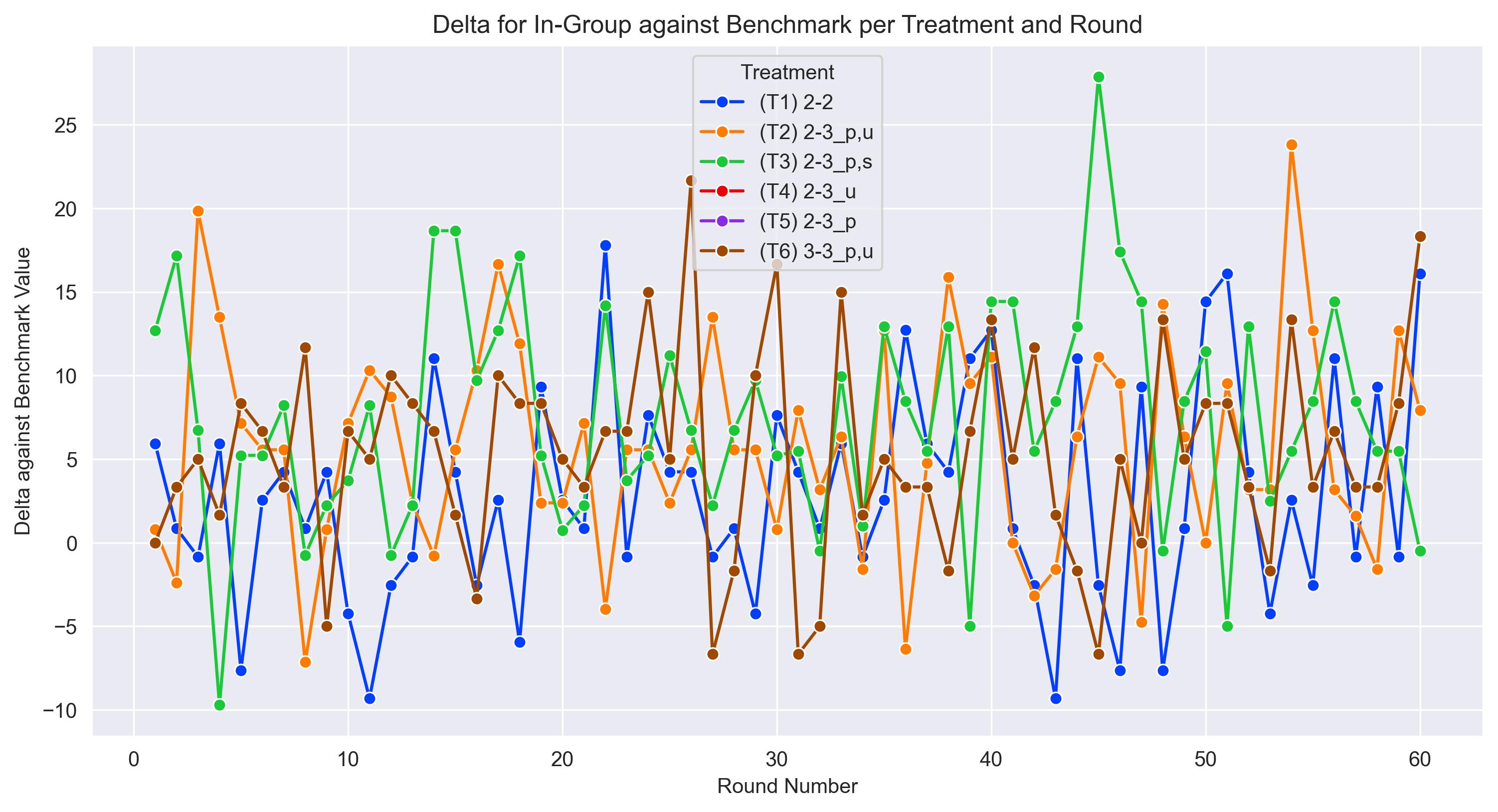}
    \caption{Average in-group deviation ($\%\Delta coins$ relative to the 0 benchmark) plotted across the 60 rounds for each treatment. Each point represents the mean $\%\Delta coins$ for the in-group in a given round. No significant time trend is observed in any treatment (OLS regression of $\%\Delta coins$ on round number yields slope coefficients not significantly different from zero in all cases)}
    \label{8-jpg}
\end{figure}

The variable is plotted in figure \ref{8-jpg}. In the regression results in table \ref{8-t1}, \ref{8-t2}, \ref{8-t3}, \ref{8-t4}, \ref{8-t5} and \ref{8-t6} we cannot detect any statistically significant effect for any treatment.

\begin{table}[ht!]
\centering

\begin{tabular}{@{}lcccccc@{}}
\toprule
                            & Coef.   & Std. Err. & t       & $P>|t|$   & [0.025   & 0.975] \\
\midrule
Intercept                   & 0.8455  & 1.694     & 0.499   & 0.620   & -2.545   & 4.236  \\
Round Number                & 0.0677  & 0.048     & 1.401   & 0.166   & -0.029   & 0.164  \\
\midrule
\multicolumn{7}{@{}l}{\textbf{Model Information:}}\\
\multicolumn{2}{l}{R-squared}       & \multicolumn{1}{r}{0.033} & \multicolumn{4}{r}{}\\
\multicolumn{2}{l}{Adj. R-squared}  & \multicolumn{1}{r}{0.016} & \multicolumn{2}{l}{No. Observations}& \multicolumn{1}{r}{60} & \multicolumn{1}{r}{}\\
\multicolumn{2}{l}{F-statistic}     & \multicolumn{1}{r}{1.963} & \multicolumn{2}{l}{Df Residuals}    & \multicolumn{1}{r}{58} & \multicolumn{1}{r}{}\\
\multicolumn{2}{l}{Prob (F-statistic)} & \multicolumn{1}{r}{0.166} & \multicolumn{2}{l}{Df Model}        & \multicolumn{1}{r}{1} & \multicolumn{1}{r}{}\\
\multicolumn{2}{l}{Log-Likelihood}  & \multicolumn{1}{r}{-196.23} & \multicolumn{2}{l}{Covariance Type} & \multicolumn{1}{r}{Nonrobust} & \multicolumn{1}{r}{}\\
\bottomrule
\end{tabular}
\caption{Time trend in in-group allocations: T1 (2-2). Tests whether \%$\Delta$coins for the in-group changes across the 60 rounds. \textit{Intercept}: estimated in-group deviation at round 0. \textit{Round Number}: change in \%$\Delta$coins per round (not significant, $p = 0.166$).}\label{8-t1}
\end{table}

\begin{table}[ht!]
\centering

\begin{tabular}{@{}lcccccc@{}}
\toprule
                            & Coef.   & Std. Err. & t       & $P>|t|$   & [0.025   & 0.975] \\
\midrule
Intercept                   & 5.2547  & 1.676     & 3.136   & 0.003   & 1.901    & 8.609  \\
Round Number                & 0.0168  & 0.048     & 0.352   & 0.726   & -0.079   & 0.112  \\
\midrule
\multicolumn{7}{@{}l}{\textbf{Model Information:}}\\
\multicolumn{2}{l}{R-squared}       & \multicolumn{1}{r}{0.002} & \multicolumn{4}{r}{}\\
\multicolumn{2}{l}{Adj. R-squared}  & \multicolumn{1}{r}{-0.015} & \multicolumn{2}{l}{No. Observations}& \multicolumn{1}{r}{60} & \multicolumn{1}{r}{}\\
\multicolumn{2}{l}{F-statistic}     & \multicolumn{1}{r}{0.1237} & \multicolumn{2}{l}{Df Residuals}    & \multicolumn{1}{r}{58} & \multicolumn{1}{r}{}\\
\multicolumn{2}{l}{Prob (F-statistic)} & \multicolumn{1}{r}{0.726} & \multicolumn{2}{l}{Df Model}        & \multicolumn{1}{r}{1} & \multicolumn{1}{r}{}\\
\multicolumn{2}{l}{Log-Likelihood}  & \multicolumn{1}{r}{-195.58} & \multicolumn{2}{l}{Covariance Type} & \multicolumn{1}{r}{Nonrobust} & \multicolumn{1}{r}{}\\
\bottomrule
\end{tabular}
\caption{Time trend in in-group allocations: T2 (2-3$_{P,U}$). Tests whether \%$\Delta$coins for the in-group changes across rounds. \textit{Round Number}: change per round (not significant, $p = 0.726$).}\label{8-t2}
\end{table}

\begin{table}[ht!]
\centering

\begin{tabular}{@{}lcccccc@{}}
\toprule
                            & Coef.   & Std. Err. & t       & $P>|t|$   & [0.025   & 0.975] \\
\midrule
Intercept                   & 6.6730  & 1.761     & 3.789   & 0.000   & 3.148    & 10.198  \\
Round Number                & 0.0292  & 0.050     & 0.581   & 0.564   & -0.071   & 0.130  \\
\midrule
\multicolumn{7}{@{}l}{\textbf{Model Information:}}\\
\multicolumn{2}{l}{R-squared}       & \multicolumn{1}{r}{0.006} & \multicolumn{4}{r}{}\\
\multicolumn{2}{l}{Adj. R-squared}  & \multicolumn{1}{r}{-0.011} & \multicolumn{2}{l}{No. Observations}& \multicolumn{1}{r}{60} & \multicolumn{1}{r}{}\\
\multicolumn{2}{l}{F-statistic}     & \multicolumn{1}{r}{0.3371} & \multicolumn{2}{l}{Df Residuals}    & \multicolumn{1}{r}{58} & \multicolumn{1}{r}{}\\
\multicolumn{2}{l}{Prob (F-statistic)} & \multicolumn{1}{r}{0.564} & \multicolumn{2}{l}{Df Model}        & \multicolumn{1}{r}{1} & \multicolumn{1}{r}{}\\
\multicolumn{2}{l}{Log-Likelihood}  & \multicolumn{1}{r}{-198.57} & \multicolumn{2}{l}{Covariance Type} & \multicolumn{1}{r}{Nonrobust} & \multicolumn{1}{r}{}\\
\bottomrule
\end{tabular}
\caption{Time trend in in-group allocations: T3 (2-3$_{P,S}$). Tests whether \%$\Delta$coins for the in-group changes across rounds. \textit{Round Number}: change per round (not significant, $p = 0.564$).}\label{8-t3}
\end{table}

\begin{table}[ht!]
\centering

\begin{tabular}{@{}lcccccc@{}}
\toprule
                            & Coef.   & Std. Err. & t       & $P>|t|$   & [0.025   & 0.975] \\
\midrule
Intercept                   &  9.0522  & 1.453     & 6.229   & 0.000   & 6.143    & 11.961  \\
Round Number                & -0.0244  & 0.041     & -0.588   & 0.559   & -0.107   & 0.059  \\
\midrule
\multicolumn{7}{@{}l}{\textbf{Model Information:}}\\
\multicolumn{2}{l}{R-squared}       & \multicolumn{1}{r}{0.006} & \multicolumn{4}{r}{}\\
\multicolumn{2}{l}{Adj. R-squared}  & \multicolumn{1}{r}{-0.011} & \multicolumn{2}{l}{No. Observations}& \multicolumn{1}{r}{60} & \multicolumn{1}{r}{}\\
\multicolumn{2}{l}{F-statistic}     & \multicolumn{1}{r}{0.3457} & \multicolumn{2}{l}{Df Residuals}    & \multicolumn{1}{r}{58} & \multicolumn{1}{r}{}\\
\multicolumn{2}{l}{Prob (F-statistic)} & \multicolumn{1}{r}{0.559} & \multicolumn{2}{l}{Df Model}        & \multicolumn{1}{r}{1} & \multicolumn{1}{r}{}\\
\multicolumn{2}{l}{Log-Likelihood}  & \multicolumn{1}{r}{-187.04} & \multicolumn{2}{l}{Covariance Type} & \multicolumn{1}{r}{Nonrobust} & \multicolumn{1}{r}{}\\
\bottomrule
\end{tabular}
\caption{Time trend in in-group allocations: T4 (2-3$_{U}$). Tests whether \%$\Delta$coins for the in-group changes across rounds. \textit{Round Number}: change per round (not significant, $p = 0.559$).}\label{8-t4}

\end{table}

\begin{table}[ht!]
\centering

\begin{tabular}{@{}lcccccc@{}}
\toprule
                            & Coef.   & Std. Err. & t       & $P>|t|$   & [0.025   & 0.975] \\
\midrule
Intercept                   & 3.5429  & 1.724     & 2.055   & 0.044   & 0.092    & 6.994  \\
Round Number                & 0.0504  & 0.049     & 1.026   & 0.309   & -0.048   & 0.149  \\
\midrule
\multicolumn{7}{@{}l}{\textbf{Model Information:}}\\
\multicolumn{2}{l}{R-squared}       & \multicolumn{1}{r}{0.018} & \multicolumn{4}{r}{}\\
\multicolumn{2}{l}{Adj. R-squared}  & \multicolumn{1}{r}{0.001} & \multicolumn{2}{l}{No. Observations}& \multicolumn{1}{r}{60} & \multicolumn{1}{r}{}\\
\multicolumn{2}{l}{F-statistic}     & \multicolumn{1}{r}{1.052} & \multicolumn{2}{l}{Df Residuals}    & \multicolumn{1}{r}{58} & \multicolumn{1}{r}{}\\
\multicolumn{2}{l}{Prob (F-statistic)} & \multicolumn{1}{r}{0.309} & \multicolumn{2}{l}{Df Model}        & \multicolumn{1}{r}{1} & \multicolumn{1}{r}{}\\
\multicolumn{2}{l}{Log-Likelihood}  & \multicolumn{1}{r}{-197.29} & \multicolumn{2}{l}{Covariance Type} & \multicolumn{1}{r}{Nonrobust} & \multicolumn{1}{r}{}\\
\bottomrule
\end{tabular}
\caption{Time trend in in-group allocations: T5 (2-3$_{P}$). Tests whether \%$\Delta$coins for the in-group changes across rounds. \textit{Round Number}: change per round (not significant, $p = 0.309$).}\label{8-t5}
\end{table}

\begin{table}[ht!]
\centering

\begin{tabular}{@{}lcccccc@{}}
\toprule
                            & Coef.   & Std. Err. & t       & $P>|t|$   & [0.025   & 0.975] \\
\midrule
Intercept                   & 4.7411  & 1.618     & 2.930   & 0.005   & 1.502    & 7.980  \\
Round Number                & 0.0203  & 0.046     & 0.441   & 0.661   & -0.072   & 0.113  \\
\midrule
\multicolumn{7}{@{}l}{\textbf{Model Information:}}\\
\multicolumn{2}{l}{R-squared}       & \multicolumn{1}{r}{0.003} & \multicolumn{4}{r}{}\\
\multicolumn{2}{l}{Adj. R-squared}  & \multicolumn{1}{r}{-0.014} & \multicolumn{2}{l}{No. Observations}& \multicolumn{1}{r}{60} & \multicolumn{1}{r}{}\\
\multicolumn{2}{l}{F-statistic}     & \multicolumn{1}{r}{0.1942} & \multicolumn{2}{l}{Df Residuals}    & \multicolumn{1}{r}{58} & \multicolumn{1}{r}{}\\
\multicolumn{2}{l}{Prob (F-statistic)} & \multicolumn{1}{r}{0.661} & \multicolumn{2}{l}{Df Model}        & \multicolumn{1}{r}{1} & \multicolumn{1}{r}{}\\
\multicolumn{2}{l}{Log-Likelihood}  & \multicolumn{1}{r}{-193.48} & \multicolumn{2}{l}{Covariance Type} & \multicolumn{1}{r}{Nonrobust} & \multicolumn{1}{r}{}\\
\bottomrule
\end{tabular}
\caption{Time trend in in-group allocations: T6 (3-3$_{P,U}$). Tests whether \%$\Delta$coins for the in-group changes across rounds. \textit{Round Number}: change per round (not significant, $p = 0.661$).}\label{8-t6}
\end{table}

\end{document}